\documentclass[a4paper,12pt]{article}
\pdfoutput=1
\usepackage{amsmath,amssymb,amsfonts,cite,color,epsf,epsfig,epstopdf,graphics,graphicx,mathtools,subcaption,physics,verbatim}
\def\thefootnote{*\arabic{footnote}}
\definecolor{ultramarine}{rgb}{0.07, 0.04, 0.56}
\definecolor{cadmiumgreen}{rgb}{0.0, 0.42, 0.24}
\definecolor{indigo(dye)}{rgb}{0.0, 0.25, 0.42}
\usepackage[linktocpage=true,breaklinks]{hyperref}
\hypersetup{
	colorlinks=true,
	citecolor=ultramarine,
	linkcolor=cadmiumgreen,
	urlcolor=indigo(dye),
}

\numberwithin{equation}{section}

\usepackage{color}

\usepackage{amsmath, amssymb, graphics, epsfig, graphicx}
\usepackage{epsf}
\usepackage{epstopdf}
\usepackage {amssymb}
\newcommand{\nc}{\newcommand}
\nc{\ba}{\begin{eqnarray}}
\nc{\ea}{\end{eqnarray}}

\newcommand{\p}{{\partial}}
\newcommand{\Mpl}{M_{\rm Pl}}
\newcommand{\A}{A^T}
\newcommand{\m}{m_{\tiny A}}
\newcommand{\f}{f}
\newcommand{\g}{h}

\newcommand{\eq}[1]{\begin{equation}#1\end{equation}}
\newcommand{\eqa}[1]{\begin{align}#1\end{align}}
\newcommand{\spl}[1]{\begin{split} #1 \end{split}}
\newcommand{\fg}[1]{\begin{figure}[tbp]\centering #1 \end{figure}}

\nc{\e}{{\bf{e}}}

\begin{document}
	
\begin{titlepage}
\setcounter{page}{1} \baselineskip=15.5pt \thispagestyle{empty}
\begin{flushright} {\footnotesize YITP-20-123, IPMU20-0103}  \end{flushright}
\vspace{5mm}
\vspace{0.5cm}
\def\thefootnote{\fnsymbol{footnote}}
\bigskip

\begin{center}
{\fontsize{22}{15}\selectfont  \bf Vector dark matter production from inflation \\ \vspace{2mm} with symmetry breaking }
\\[0.5cm]
\end{center}
\vspace{0.2cm}

\begin{center}
{ Borna Salehian$^{1}\footnote{salehian@ipm.ir}$, Mohammad Ali Gorji$^{2}\footnote{gorji@yukawa.kyoto-u.ac.jp}$, Hassan Firouzjahi$^{1}\footnote{firouz@ipm.ir}$, 
Shinji Mukohyama$^{2,3}\footnote{shinji.mukohyama@yukawa.kyoto-u.ac.jp}$  }
\\[.7cm]

{\small \textit{$^{1}$School of Astronomy, 
Institute for Research in Fundamental Sciences (IPM) \\ 
 P.~O.~Box 19395-5531, Tehran, Iran }} \\

 {\small \textit{$^{2}$Center for Gravitational Physics,
 Yukawa Institute for Theoretical Physics \\
 Kyoto University, 606-8502, Kyoto, Japan}} \\

 {\small \textit{$^{3}$Kavli Institute for the Physics and Mathematics of the Universe (WPI), 
 The University of Tokyo Institutes for Advanced Study, 
 The University of Tokyo, Kashiwa, Chiba 277-8583, Japan}} \\

\end{center}

\vspace{1cm}


\hrule \vspace{0.5cm}
{\small  \noindent \textbf{Abstract} \\[0.2cm]
\noindent
We present a scenario of vector dark matter production from symmetry breaking at the end of inflation. In this model, the accumulated energy density associated with  the quantum fluctuations of the dark photon accounts for the present energy density of dark matter. The inflaton is a real scalar field while a heavy complex scalar field, such as the waterfall of hybrid inflation, is charged under the dark gauge field. After the heavy field becomes tachyonic  at the end of inflation,  rolling rapidly  towards its global minimum,  the dark photon acquires mass via the Higgs mechanism. To prevent the decay of the vector field energy density during inflation, we introduce couplings between the inflaton and the gauge field such that the energy is pumped to the dark sector.  The setup can generate the observed dark matter abundance for a wide range of the dark photon's mass and with the reheat temperature around $10^{12}$ GeV. The model predicts the formation of cosmic strings at the end of inflation with the tensions which are consistent with the CMB upper bounds. 
\vspace{0.5cm} \hrule}

\end{titlepage}

\section{Introduction}
\label{sec:intro}

The nature of dark matter is still unknown and remains one of the most compelling evidences for the physics beyond the standard model of particle physics (SM). While the most well-known candidate are the so-called Weakly Interacting Massive Particles (WIMPs), in the absence of observation of such new particles in colliders the researchers tend to look for other possibilities as well (see Refs.~\cite{Aprile:2017aty,Aprile:2020tmw,Farzan:2020dds,Shakeri:2020wvk,Cai:2020kfq} for attempts in this regard). Recently, dark photon became one of the candidate dubbed ``vector dark matter'' \cite{Essig:2013lka,Nelson:2011sf,Arias:2012az,Graham:2015rva, Kolb:2020fwh}. For a recent review on dark photons see Ref.~\cite{Fabbrichesi:2020wbt}. They are supposed to have negligible direct interactions with the SM particles while they can have significant direct coupling with other particles in the dark sector as well as the universal minimal coupling to gravity. It is then shown that they can be efficiently produced before the time of matter radiation equality and play the role of dark matter relic density \cite{Agrawal:2017eqm,Co:2017mop,Agrawal:2018vin,Co:2018lka,Dror:2018pdh}. While most experimental efforts for dark photon searches rely on its weak coupling to the SM particles, they can be probed indirectly through the gravitational waves observations and this possibility has been the subject of many studies in recent years \cite{Pierce:2018xmy,Machado:2018nqk,Machado:2019xuc,Salehian:2020dsf,Michimura:2020vxn,Namba:2020kij}. 

On the other hand inflation has emerged as the leading paradigm to solve  the horizon and the flatness problems associated with the big bang cosmology and also to explain the origin of the large scale structures in the Universe. Among the basic predictions of simple models of inflation are that the primordial perturbations are nearly scale invariant, nearly Gaussian and nearly adiabatic which are well consistent with the cosmological observations \cite{Akrami:2018odb, Akrami:2019izv}. The inflaton field  is usually taken to be a scalar field which slowly rolls on top of a nearly flat  potential leading to a quasi-de Sitter background expansion.  However, there is no primary reason which forbids considering other fields like gauge bosons and even fermions during inflation. Evidently, it is not an easy task to find a consistent inflationary dynamics completely based on the fermionic fields while inflationary models based on the gauge fields are widely studied in recent years. Indeed, it is possible to find an inflationary scenario completely based on vector/gauge fields \cite{Golovnev:2008cf,Maleknejad:2011jw} or considering scenarios in which gauge fields partially contribute to the dynamics of inflation. One problem associated to the presence of the gauge fields during inflation is the conformal symmetry which causes the  exponential decay of the energy density of the gauge fields in a quasi-de Sitter background. One natural possibility to remedy this issue is to consider a  direct coupling between the inflaton and the vector/gauge fields. In this respect the gauge field drags energy from the inflaton which can prevent the decay of its energy density. There are some models based on this picture among which are the so-called anisotropic inflation \cite{Watanabe:2009ct, Watanabe:2010fh, Bartolo:2012sd, Emami:2010rm, Emami:2013bk} and pseudoscalar inflaton \cite{Anber:2006xt,Sorbo:2011rz, Barnaby:2010vf, Barnaby:2011qe}. This idea was already implemented to explain the origin of the primordial magnetic field \cite{Ratra:1991bn,Garretson:1992vt,Anber:2006xt, Martin:2007ue, Demozzi:2009fu,Kanno:2009ei,Emami:2009vd,Fujita:2012rb,Caprini:2014mja,Caprini:2017vnn, Schober:2020ogz, Talebian:2020drj}.

Dark matter and inflation are usually considered to be unrelated. However, inflation 
and reheating  can be the  frameworks for the productions of particles beyond the SM. As the dark matter particles may be from beyond the SM sector, it is interesting to look for the possibility of the production of dark matter particles during inflation. In this regard, it was shown that if one considers a massive dark gauge field during inflation without any direct coupling to inflaton, the associated longitudinal mode can be responsible for the dark matter energy density after the time of matter and radiation equality \cite{Graham:2015rva}. Very recently, this idea was further extended and it was shown that even the transverse modes can be responsible for the dark matter relic if a massive gauge boson directly interacts with the inflaton \cite{Bastero-Gil:2018uel,Nakayama:2019rhg,Nakayama:2020rka,Nakai:2020cfw}. The difficulty with the dark matter models based on the transverse modes of the gauge fields produced during inflation is that on one hand, the gauge field should have a small mass compared to the energy scale of inflation to have efficient dark photon productions during inflation. While, on the other hand, the mass should be large enough to make the produced dark photons non-relativistic before the time of matter and radiation equality. These criteria make the appropriate mass range of the dark photon to be very model-dependent.  For example  in Ref. \cite{Bastero-Gil:2018uel} with the parity violating interaction between the inflaton and the gauge field the mass of dark photons should be $\m \geq 10^{-6}$ eV while in a model with non-minimal conformal coupling one finds $\m \geq 10^{-21}$ eV \cite{Nakai:2020cfw}.

In this paper, we consider a scenario in which the dark gauge field acquires mass dynamically through a symmetry breaking, i.e. the Higgs mechanism, at the end of inflation. Apart from the fact that the Higgs mechanism is a natural way of giving mass to a gauge boson in the SM or beyond SM, it opens up a new window to the scenarios of vector dark matter production during inflation. The reason is that the gauge field is massless during inflation and therefore dark photons can be produced very efficiently. The mechanism of  dark photon production that we consider here is based on particle production through the non-adiabatic evolution of the mode functions of the dark photon from the direct coupling between the gauge field and inflaton. This is different from the more popular thermal particle production scenarios, i.e. freeze-out or -in \cite{Hall:2009bx}, and also the so-called non-thermal misalignment mechanism \cite{Nelson:2011sf} in the context of dark matter while essentially the same mechanism is commonly considered for producing primordial metric fluctuations observed in the cosmic microwave background (CMB). Thanks to the Higgs mechanism, the gauge field then acquires mass only toward the end of inflation and one can easily find appropriate range of parameters for which the produced dark photons can become non-relativistic before the time of matter-radiation equality and also to give the correct dark matter abundance. In this model the mass of dark photon can easily 
be so large that they become non-relativistic even right after inflation. In this respect, we will be able to find new novel mechanisms which were not possible in previous works  due to the assumption of small mass for the dark photons during inflation. 

The rest of the paper is organized as follows. In Section \ref{sec-SB} we present a general inflationary model in which a dark gauge boson acquires mass through the symmetry breaking at the end of inflation. In Section \ref{sec-DP}, we show that dark photon can be efficiently produced during inflation and they can play the role of (vector) dark matter after inflation. We show that the symmetry breaking scenario opens a large parameter space for these models. In Section \ref{sec-Model} we employ our general scenario to a particular case and show that it has a rich phenomenology. In Sections \ref{sec-model-kinetic} and \ref{sec-model-CS} we consider two subsets of this specified model which resemble the already well-known models of kinetic conformal coupling and the parity violating coupling respectively. In Section \ref{sec-model-general} we then look at the more general case which includes both of the conformal and parity violating couplings. Section \ref{summary} is devoted to the summary and conclusions.

\section{Inflation with symmetry breaking}\label{sec-SB}

In this section we present an inflationary model in which a gauge boson in the beyond SM dark sector (dark photon) obtains mass through the dynamical symmetry breaking at the end of inflation. The symmetry breaking mechanism can be achieved via a waterfall phase transition as in hybrid inflation \cite{Linde:1993cn,Copeland:1994vg}. Note that the details of the inflationary model is not important for our computation of the dark photon evolution and its relic energy density. The only important aspect of the model is that the symmetry breaking occurs approximately at the end of inflation such that the gauge field is massless during inflation but massive afterwards. Having said this, it is possible to consider a model in which the symmetry breaking takes place  during inflation but we leave this possibility to  a future study.

The standard scenario where the symmetry is broken at the end of inflation is the hybrid inflation in which the inflaton is a real scalar field and there is another complex scalar field, the waterfall field, which is responsible for the symmetry breaking and the termination of inflation. We extend this standard picture by assuming that the waterfall field is charged under the dark photon  field.  The theory is described by the following action \cite{Emami:2011yi,Abolhasani:2013bpa}
\begin{eqnarray}
\label{action} S = \int
d^4 x  \sqrt{-g} \bigg[ \frac{\Mpl^2}{2} \mathcal{R} - \frac{1}{2}(\p\phi)^2
- \frac{1}{2} |D\psi|^2 - V(\phi,|\psi|) 
- \frac{1}{4} \f^{2}(\phi) F_{\mu \nu} F^{\mu \nu} - \frac{1}{4} \g^{2}(\phi) F_{\mu \nu} {\tilde F^{\mu \nu}} \bigg] \,,
\end{eqnarray}
where $\Mpl = 1/ \sqrt{8 \pi G}$ is the reduced Planck mass with $G$ being the Newton constant and $\mathcal{R}$ is the Ricci scalar.  As mentioned before, the inflaton is the real scalar field $\phi$ and the waterfall is the complex scalar field $\psi$ which is charged under the $U(1)$ gauge field $A_\mu$. The components of the covariant derivative of the waterfall field is given by
\ba
D_\mu
\psi = \partial_\mu  \psi + i \e \,  \psi  \, A_\mu \,,
\ea
where $\e$ is the dimensionless gauge coupling between the gauge field $A_\mu$ and $\psi$. As usual, the components of the field strength tensor $F$ associated to the gauge field is given by
\ba F_{\sigma \rho} = \nabla_\sigma A_\rho
- \nabla_\rho A_\sigma  = \partial_\sigma A_\rho - \partial_\rho A_\sigma \, ,
\ea
and $\tilde{F}^{\sigma\rho}=\epsilon^{\sigma\rho\delta\eta}F_{\delta\eta}/2$ is the corresponding dual field strength tensor. 

We have added the conformal couplings $\f(\phi)$ and $\g(\phi)$ to break the conformal invariance and to pump energy from the inflaton sector to the gauge field sector. Indeed, it is well known that in the simple Maxwell theory with no such couplings, the electromagnetic fields energy density are exponentially diluted during the quasi-de Sitter inflationary expansion. To prevent this, one usually introduces couplings via appropriate forms of $\f(\phi)$ and $\g(\phi)$. This formalism was employed extensively in the context of anisotropic inflation where it is shown that with  appropriate functional form for $\f(\phi)$ with $\g(\phi)=0$, the energy density of the gauge field survives the de Sitter expansion while its perturbations become nearly scale invariant \cite{Watanabe:2009ct, Watanabe:2010fh, Bartolo:2012sd, Emami:2010rm, Emami:2013bk}. For a review of anisotropic inflation see Ref. \cite{Emami:2015qjl}. On the other hand, in the pseudo-scalar inflation with the parity violating coupling $\g(\phi)^2 \propto \phi$ and $\f(\phi)=1$, vector field perturbations are enhanced through a tachyonic instability \cite{Anber:2006xt,Sorbo:2011rz, Barnaby:2010vf, Barnaby:2011qe}. Both of these scenarios are also implemented as a mechanism for generating primordial magnetic field during inflation 
\cite{Ratra:1991bn,Garretson:1992vt,Anber:2006xt, Martin:2007ue, Demozzi:2009fu,Kanno:2009ei,Emami:2009vd,Fujita:2012rb,Caprini:2014mja,Caprini:2017vnn, Schober:2020ogz, Talebian:2020drj}.   We will study the production of dark photon during inflation in the following section. Here, we describe the inflationary model.

We assume the axial symmetry so the potential depends on the modulus of the waterfall field $|\psi|$ along with the inflaton field. We thus decompose the waterfall field into the radial and angular components as follows
\ba
\label{phi}
\psi \equiv |\psi| e^{i \theta}\,,
\ea
where the field $\theta$ determines the argument of the complex waterfall field. As usual, the action
(\ref{action}) is invariant under the local gauge transformation \ba
\label{transformation} A_\mu \rightarrow A_\mu - \frac{1}{\e}
\partial_\mu \epsilon(x) \quad , \quad \theta \rightarrow \theta +
\epsilon(x) \,. 
\ea
As a result, we can exploit this gauge symmetry to set $\theta=0$, i.e. the unitary gauge. In this gauge, we deal with a real waterfall field in practice and, therefore, from now on we do not write the absolute value sign anymore. The typical choice for the potential in hybrid inflation is \cite{Linde:1993cn,Copeland:1994vg}
\begin{equation}
\label{pot}
V(\phi,\psi) = \frac{\lambda}{4} \left( \psi^2 - \frac{M^2}{\lambda}\right)^2 
+ \frac{1}{2} m^2 \phi^2 + \frac{1}{2} {g}^2 \phi^2\psi^2 \, ,
\end{equation}
where $\lambda$ and $g$ are two dimensionless couplings. The potential is depicted in Fig.~\ref{fig:hybrid}. 
\fg{
\centering
\includegraphics[width=0.55\textwidth]{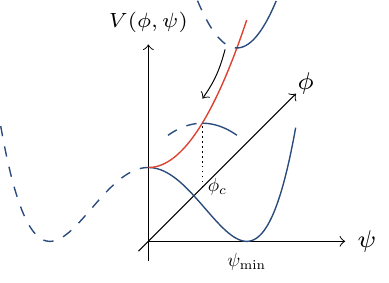}
\caption{A schematic view of symmetry breaking based on hybrid inflation.  The red line is the chaotic potential for $\phi$ while the blue curve is the potential of $\psi$ which is initially very heavy but at $\phi_c$ becomes tachyonic and soon settles at its global minimum at $\psi_{\rm min}$. Note that we always have $\psi\geq0$.}
\label{fig:hybrid}
}
Initially the inflaton starts in the region $\phi>\phi_c$. Here, $\phi_c \equiv M/g$ is a critical value beyond which the waterfall field $\psi$ has positive mass squared. The assumption is that $\psi$ is very heavy so during inflation and it is stuck in its local minimum $\psi=0$ and one can safely ignore its quantum fluctuations. We assume that inflation is mainly driven by the vacuum term $M^4/4 \lambda$. In other words the potential during inflation is
\eq{\label{V}
V(\phi,0) \approx\frac{1}{2}m^2\phi^2+\frac{M^4}{4\lambda}\,,\qquad \frac{M^4}{4 \lambda} \gg \frac{1}{2} m^2 \phi^2\,.
}
This is crucial because we need inflation to end after the symmetry breaking and it requires that the waterfall dominates the energy density during inflation. As the inflaton field rolls during the expansion of the Universe and reaches the critical value $\phi_c$, the waterfall field becomes tachyonic and rapidly rolls towards its global minimum at $\psi_{\rm min} =  M/\sqrt \lambda$ (note that $\psi$ is the absolute value of the waterfall field so it can only be positive) terminating inflation efficiently.  We define the following dimensionless parameters 
\ba
\label{alpha-beta}
\alpha \equiv \frac{m^2}{H^2}\,, \hspace{1cm}
\beta \equiv \frac{M^2}{H^2}\,,
\ea
which measure the ratios of the mass scales of the inflaton and the waterfall field to the  Hubble parameter $H$ during inflation. The condition that $\phi$ is slow-rolling implies that $\alpha \ll 1$ while the assumption that $\psi$ is very heavy during inflation such that we can ignore its quantum fluctuations, requires $\beta \gg1$. Further, we want the procedure of phase transition happens very rapidly within less than one e-fold of inflation which implies $\alpha\beta\gg1$ \cite{Abolhasani:2010kr,Abolhasani:2011yp, Abolhasani:2012px}.  

Since during inflation the waterfall  is stuck in its local minimum  $\psi=0$, the induced mass of the dark photon field is zero during inflation, $\m=0$. As a result, the dark photon mode function receives no corrections from the mass term and it is given by the conventional massless modes in a nearly fixed de Sitter background. As inflation ends and the waterfall field rapidly settles to its global minimum, the induced mass of the dark photon is given by
\ba
\label{mA}
\m^2 =  \frac{\e^2 M^2}{\lambda}\,,\qquad\qquad(\text{at the end of inflation})\,.
\ea

We also define the ratio of the gauge field mass to the Hubble scale at the end of inflation $H_{\rm e}$ as follows
\eq{
\label{R}
R\equiv\frac{\m^2}{H_{\rm e}^2}\simeq\frac{\e^2}{\lambda}\beta. 
}
Depending on the three parameters $\e$, $\lambda$ and $\beta$ we can have dark photon heavier ($R>1$) or lighter ($R<1$) than the Hubble scale at the end of inflation $H_{\rm e}$.

The above scenario was a working example of symmetry breaking at the end of inflation to yield mass to the dark photon via the Higgs mechanism. The specific setup discussed above  is based on hybrid model but we emphasize that this picture is more general and not restricted to the particular setup of hybrid inflation. Indeed, the standard hybrid model is ruled out from the CMB observations as it generates a blue spectral index for the curvature perturbation power spectrum \cite{Bassett:2005xm}. Instead, all we need is a phase of inflation with a nearly constant value of $H$ terminated by a heavy waterfall field which is not excited during inflation. There are various ways that the standard hybrid setup can be modified such that the setup will be consistent with the CMB observations. One simple extension is to consider a multiple field models where more than two fields can drive inflation and generate the large scale perturbations. One can massage the inflationary potential such that on the observed CMB scales the spectrum is red-tilted while towards the final stage of inflation the model effectively looks like Eq. (\ref{pot}) with one field essentially playing the role of the inflaton field. Alternatively, one may embed our setup in the context of brane inflation \cite{Dvali:1998pa, Kachru:2003sx} in string theory where the inflaton field is the geometric distance between a stack of branes and that of anti-branes in the string theory compactification. As the branes and anti-branes approach each other a tachyon develops and inflation terminates rapidly via the branes and anti-branes annihilation. One may think of the dark photon as living in some left over brane in these configuration which is different than our three-brane where the SM particles are located. 

A generic prediction of our setup is that cosmic strings are produced at the end of inflation. This is a general outcome when a $U(1)$ symmetry is broken in 
early universe \cite{Kibble:1976sj, Vilenkin:2000jqa}. Unlike monopoles and domain walls, cosmic strings are allowed to form in cosmic history with interesting cosmological implications such as lensing, CMB anisotropies or generating stochastic gravitational waves \cite{Vilenkin:2000jqa}.  The tension of string $\mu$ is related to the scale of symmetry breaking 
via  \cite{Vilenkin:2000jqa}, 
\ba
\label{string-mu}
\mu \sim \frac{M^2}{ \lambda} \, .
\ea
The CMB observations  impose the upper bound $G \mu \lesssim 10^{-7}$ \cite{Ade:2013xla}. We will use this upper bound on $\mu$ to impose some bounds on our model parameters such as the value of  $R$ and the reheat temperature.  

In the following sections we will focus on the production of dark photons during inflation and their evolution afterwards 
to form dark matter observed today.

\section{Dark photons: Production and evolution}\label{sec-DP}

As we explained in the previous section, around the time of end of inflation, the dark gauge field acquires mass from the waterfall field. Thus, the dark photon sector of the action \eqref{action} takes the following form
\ba
\label{actionA} S_{\tiny A}= \int
d^4 x  \sqrt{-g} \left [ - \frac{1}{4} \f^{2}(\phi) F_{\sigma\rho} F^{\sigma\rho} 
- \frac{1}{4} \g^{2}(\phi) F_{\sigma\rho} \tilde{F}^{\sigma\rho} 
- \frac{1}{2}\m^2 \, A_\sigma A^\sigma \right]\,.
\ea
Note that the mass term in the above action is generated dynamically through the symmetry breaking scenario that we presented in the previous section. Therefore, during inflation dark photons are massless $\m=0$ while at the end of inflation they almost instantaneously  acquire mass $\m\neq0$. This is the key feature of our model which makes it completely different from other vector dark matter models which are based on the inflationary scenarios.

Taking variation of the action \eqref{actionA} with respect to $A_{\sigma}$, we find the equation of motion for the gauge field as
\begin{equation}\label{Maxwell}
\nabla^\rho \big( \f^2 F_{\rho\sigma} + \g^2 {\tilde F}_{\rho\sigma} \big) - \m^2 A_\sigma = 0 \,.
\end{equation}
The energy momentum tensor of the gauge field can also be obtained by taking the variation of  Eq.~\eqref{actionA} with respect to the metric, yielding
\eq{\label{EMT-EM}
	T_{\sigma\rho}= \f^2F_{\sigma\eta}F_\rho{}^\eta-\frac{1}{4}\f^2g_{\sigma\rho}F_{\eta\delta}F^{\eta\delta}
	+\m^2\Big(A_\sigma A_\rho-\frac{1}{2}g_{\sigma\rho}A_\eta A^\eta\Big)\,.
}
We consider a spatially flat  Friedmann-Lema\^{i}tre-Robertson-Walker (FLRW) metric for the inflationary background geometry
\begin{align}\label{FLRW}
{\rm d}s^2 = a^2(\tau) \big(-{\rm d}\tau^2+\delta_{ij}~{\rm d}x^i{\rm d}x^j \big) \, ,
\end{align}
in which $a(\tau)$ is the scale factor and $\tau$ is the conformal time related to the cosmic time via $d\tau = dt/a(t)$. 

For the gauge field sector, our assumption is that the gauge field has no background values $\ev{A_\sigma}=0$. Taking this fact into account, from the temporal component of Eq.~\eqref{Maxwell} we find
\eq{\label{Eq.long}
\big(\f^2\nabla^2-\m^2a^2\big)A_0=\f^2(\p_iA_i)'\,,
}
where a prime denotes derivative with respect to the conformal time. Therefore,  as usual, the component $A_0$ is a constraint and its solution can be imposed at the level of the action. Considering the background symmetry we decompose the spatial components of the gauge field as
\eq{\label{dec-A}
A_i=\p_i\chi+\A_i\,,\hspace{1cm}\p_i\A_i=0\,,
}
where $\chi$ and $\A_i$ are longitudinal and transverse parts of the gauge field respectively. Substituting the above relation in Eq.~\eqref{Eq.long}, we can solve $A_0$ in favour of the longitudinal mode $\chi$ as follows
\eq{
	\label{A0}
A_0=\Big(\frac{\f^2\nabla^2}{\f^2\nabla^2-\m^2a^2}\Big)\chi'\,.
}

The energy density of the gauge field is defined by $\rho_{\tiny A}\equiv-T^0{}_0$. We decompose it into  $\rho_{\tiny A}=\rho_T+\rho_L+\p_iJ_i$ in which $\rho_T$ and $\rho_L$ are the energy densities for the transverse and longitudinal modes respectively. The term $\p_iJ_i$ is a total divergence term which after taking the vacuum expectation value does not contribute to the background \cite{Nakai:2020cfw}. Now,  plugging Eq. \eqref{A0} in Eq.  (\ref{EMT-EM}) and after a couple of lines of algebra, we find
\eq{
\label{rhoT}
\rho_T=\frac{\f^2}{2a^4}\left[(\A_i{}')^2+(\p_i\A_j)^2+\frac{\m^2a^2}{\f^2}(\A_i)^2\right]\,,\quad
\rho_L=\frac{\m^2}{2a^2}\left[\chi'\big(\frac{\f^2\nabla^2}{\f^2\nabla^2-\m^2a^2}\big)\chi'+(\p_i\chi)^2\right]\,.
} 

Substituting  Eq.~\eqref{dec-A} into the action (\ref{actionA}) and then eliminating $A_0$ in favour of the longitudinal mode $\chi$ through Eq.~\eqref{A0}, we can also decompose the gauge field quadratic action into the transverse and longitudinal parts as follows
\eq{
	 S_{\tiny A} = S_T+S_L\,,
}
where we have defined
\eqa{\label{ST}
	S_T&=\frac{1}{2}\int\dd[3]{x}\dd{\tau}\Big[ \f^2 (\A_i{}')^2
	- \f^2 (\p_i\A_j)^2-\m^2a^2(\A_i)^2 + 2 \g\g' \epsilon_{ijk} \A_i \p_j \A_k \Big] \,, \\
	\label{SL}
	S_L&=\frac{\m^2}{2}\int\dd[3]{x}\dd{\tau}a^2\left[\chi'\big(\frac{\f^2\nabla^2}{\f^2\nabla^2-\m^2a^2}\big)\chi'-(\p_i\chi)^2\right] \,.
}
During inflation the gauge field is massless $\m=0$ and therefore the longitudinal mode is absent. After the symmetry breaking we have $\m\neq0$ and the longitudinal mode is dynamical with a large mass. As the conformal and parity violating couplings are stabilized after inflation the massive longitudinal mode cannot be excited efficiently. As a result, from now on we neglect the longitudinal mode and focus only on the transverse modes.

\subsection{Quantization} \label{sec:quant}

To study the production of dark photons, we need to look at the quantum fluctuations of the transverse modes of the dark gauge field during inflation. We define the canonical field $V_i\equiv f\A_i$ and expand it in terms of the creation  and annihilation operators $a_{\vb{k}}$ and $a^\dagger_{\vb{k}}$  as usual
\begin{equation}\label{V-mode}
V_i=\sum_\lambda\int\frac{\dd[3]{k}}{(2\pi)^{3/2}} \varepsilon^\lambda_i({\bf k})
\big[v_{k,\lambda}(\tau)a_{\vb{k},{\lambda}}
+v_{k,\lambda}(\tau)^*a^\dagger_{-\vb{k},{\lambda}} \big] e^{i\vb{k}.\vb{x}}\,, \hspace{1cm} 
[a_{\vb{k},{\lambda}}, a^\dagger_{\vb{k}',{\lambda'}}] = \delta_{\lambda\lambda'}\delta(\vb{k}-\vb{k}')\,,
\end{equation}
where the time dependence of the gauge field is encoded in the mode functions $v_{k,\lambda}(\tau)$. In the above relation, $\varepsilon^\lambda_i({\bf k})$ are the polarization vectors for $\lambda =\pm$ which satisfy $\varepsilon_i^{\lambda}({\bf k}){}^*=\varepsilon_i^{-\lambda}({\bf k})=\varepsilon_i^{\lambda}(-{\bf k})$, $k_i\varepsilon_i^\lambda({\bf k})=0$, and also the identity $\epsilon_{ij\ell}k_j\varepsilon_\ell^\lambda({\bf k})=-i\lambda{k}\varepsilon_i^\lambda({\bf k})$. (see appendix A of Ref. \cite{Salehian:2020dsf} for more details).

During inflation, the field is massless $\m=0$ so substituting from Eq.~\eqref{V-mode} in the action \eqref{ST}, and taking the variation, the equation of mode function is given by 
\ba\label{MF-Eq}
v_{k,\lambda}'' + \Big( k^2 - 2 \lambda k \frac{\g\g'}{\f^2} -\frac{\f''}{\f} \Big) v_{k,\lambda} =0   \, .
\ea
Note that for nonzero time-dependent coupling $\g$ the mode functions for different polarizations evolve differently. 

The energy density of the gauge field during inflation can be computed from Eq.~\eqref{rhoT}
\eq{
	\label{rhoT3}
	\rho_{\tiny A}\simeq \rho_T=\frac{1}{2a^4} \sum_{\lambda} \int\frac{\dd[3]{k}}{(2\pi)^3}
	\Big[ \big|{ \f\big(\frac{v_{k,\lambda}}{\f}\big)'} \big|^2+k^2\abs{v_{k,\lambda}}^2\Big]\,,\qquad\text{(during inflation)}\,.
}  
We will use this expression in computing the backreaction of the vector field on inflationary background dynamics. However, at the end of inflation the dark photon becomes massive through the symmetry breaking process. As a result, after that the dark photon  is a massive vector field with a mode function denoted by $u_{k,\lambda}$ and its equation of motion reads
\ba\label{MF-Eq2}
u_{k,\lambda}'' + \Big( k^2 - 2 \lambda k \frac{\g\g'}{\f^2} 
-\frac{\f''}{\f} + \frac{\m^2a^2}{\f^2} \Big) u_{k,\lambda} =0\,,
\qquad\text{(after symmetry breaking)}\,.
\ea
Note that the symmetry breaking occurs at the end of inflation so the above equation is relevant only around the time $\tau=\tau_{\rm e}$. 

We want to compute the energy density of the dark photons after acquiring mass at the end of inflation which from Eq.~\eqref{rhoT} is given by
\eq{
	\label{rhoT4}
	\rho_{\rm e}=\frac{1}{2a_{\rm e}^4} \sum_{\lambda} \int\frac{\dd[3]{k}}{(2\pi)^3}
	\Big[ \big|{\f\big(\frac{u_{k,\lambda}}{\f}\big)'}\big|^2+\big(k^2+\frac{\m^2a^2}{\f^2}\big)\abs{u_{k,\lambda}}^2\Big]\bigg|_{\tau_{\rm e}}\,. 
} 
This will be carried to the radiation and matter dominated eras as the observed relic abundance of the dark matter. To solve Eq.~\eqref{MF-Eq2} exactly, we need to take into account the dynamics of the waterfall field during and after the phase transition (see Refs. \cite{Abolhasani:2010kr,Gong:2010zf}). However, since we have assumed that the waterfall field is heavy enough, the symmetry breaking happens almost instantaneously around the time of end of inflation $\tau_{\rm e}$  \cite{Linde:1993cn,Abolhasani:2010kr}, and we can avoid the technicalities of considering the waterfall field dynamics. 

Assuming an instantaneous phase transition, we only need to impose the junction conditions $u_{k,\lambda}(\tau_{\rm e})=v_{k,\lambda}(\tau_{\rm e})$ and $u'_{k,\lambda}(\tau_{\rm e})=v'_{k,\lambda}(\tau_{\rm e})$ which enables us to find the energy density of the dark photons at the end of inflation in terms of the massless mode function $v_{k,\lambda}$ as follows
\eq{
	\label{rhoT5}
	\rho_{\rm e}=\frac{1}{2a_{\rm e}^4} \sum_{\lambda} \int\frac{\dd[3]{k}}{(2\pi)^3}
	\left[ \big| {\f\big(\frac{v_{k,\lambda}}{\f}\big)'}\big|^2+\big(k^2+\frac{\m^2a^2}{\f^2}\big)\abs{v_{k,\lambda}}^2\right] \bigg|_{\tau_{\rm e}}\,,\,\text{(at the end of inflation)}\,.
} 
As we will see in the concrete models in the following, we can parametrize the energy density of the produced dark photons at the end of inflation as 
\begin{equation}\label{rho-param}
\rho_{\rm e} \equiv \mathcal{C}\, H_{\rm e}^4 \,,
\end{equation}
where $H_{\rm e}$ is the Hubble parameter at the end of inflation and 
$\mathcal{C}$ is a dimensionless function which depends on the parameters of the system under consideration. We will find explicit form of $\mathcal{C}$ for some particular cases in the next sections.

In the following subsections we use expressions \eqref{rhoT3} and \eqref{rhoT5} for the backreaction constraint and the relic abundance of the dark matter respectively. We remind that  Eq. \eqref{rhoT3} is the energy density accumulated from the gauge field fluctuations  during inflation while $\rho_{\rm e}$
 in Eq. \eqref{rhoT5}  is the energy accumulated at the end of inflation including the effects of mass term $\m^2$. The energy density of the dark photon during the big bang expansion is related to $\rho_{\rm e}$ via $\rho_{\tiny A} \propto \rho_{\rm e}/a^4$ or $\rho_{\tiny A} \propto \rho_{\rm e}/a^3$, depending on whether the dark photon is relativistic or non-relativistic. Note that here we are neglecting any form of change in the energy density of dark photons after inflation other than the expansion of the universe. Indeed, the energy density of dark photon can vary due to the coupling to SM or dark sector particles by decaying into light particles or interacting with the thermal bath during the radiation dominated epoch. While the detailed study of these effects are beyond the scope of this paper, we briefly discuss these issues in the following subsections.

\subsection{Constraints on the model}

Our assumption is that the gauge field does not destroy the coupled equations of motion for the inflaton and the background geometry. In this respect, there are two sources of backreaction from dark photons to the dynamics of the inflaton and the geometry. Correspondingly, there are two conditions to be imposed. The first condition is that the energy density of dark photons must be small compared to the vacuum energy during inflation, i.e.
\eq{\label{BR-1}
\rho_{\tiny A}\ll 3\Mpl^2H^2\,.
}
Second, the source term that appears in the Klein-Gordon equation due to the conformal couplings must be small. The equation of motion for the inflaton reads
\begin{equation}\label{KG}
\Box\phi - \dv{V}{\phi} = S_\phi \,, \hspace{1cm} 
S_\phi \equiv \frac{1}{2}\f\dv{\f}{\phi}F_{\mu\nu}F^{\mu\nu}
+ \frac{1}{2}\g\dv{\g}{\phi}F_{\mu\nu}{\tilde F}^{\mu\nu} \,.
\end{equation}
Note that the right hand side of the Klein-Gordon equation is nonzero because of the couplings $\f(\phi)$ and $\g(\phi)$ between the inflaton and the gauge field as given in the action of gauge field Eq.~\eqref{actionA}. In the FLRW background \eqref{FLRW}, the equation of motion for the inflaton \eqref{KG} takes the form
\eq{\label{S}
\ddot{\phi}+3H\dot{\phi}+\dv{V}{\phi}=-S_\phi\,,\qquad S_\phi\equiv 
\frac{-1}{a\dot{\phi}} \Big[
\f\f' (E_iE_i-B_iB_i) - 2 \g\g' E_i B_i
 \Big] \,,
}
where $E_i = - a^{-2} (A'_i - \p_i A_0)$ and $B_i = a^{-2} \epsilon_{ijk} \p_j A_k$ are the corresponding electric and magnetic fields of the dark gauge field and we have used $d\f/d\phi=\f'/a\dot{\phi}$ with a dot denoting the  derivative with respect to the cosmic time $t$. The second backreaction condition then is given by
\eq{\label{BR-2}
|S_\phi| \ll 3H \dot{\phi} \,.
}
For the later purposes let us write $S_\phi$ given in Eq.~\eqref{S} in terms of the mode functions
\eq{
	\label{SS}
	S_\phi=\frac{-1}{a^5\dot{\phi}}\sum_\lambda\int\frac{\dd[3]{k}}{(2\pi)^3}\left\{\frac{\f'}{\f}\Big(\Big|{\f\Big(\frac{v_{k,\lambda}}{f}\Big)'}\Big|^2-k^2\abs{v_{k,\lambda}}^2\Big)+2\lambda k\frac{\g\g'}{f^2}\Re \Big[ v_{k,\lambda}^*\f\big(\frac{v_{k,\lambda}}{f}\big)' \Big] \right\}\,.
}

The conditions \eqref{BR-1} and \eqref{BR-2} should be satisfied in order to have a consistent inflationary background. Note that both of the above conditions must always be valid during inflation. However, since the gauge field energy density and also $S_\phi$ are growing, the two conditions are strongest at the end of inflation. Thus we will evaluate them at the end of inflation.

Moreover, we should look at the effects of the gauge field perturbations on the curvature perturbations as well. We need to demand that the corrections in power spectrum, bispectrum and trispectrum  induced  from the gauge field perturbations are suppressed with respect to the standard curvature perturbations given by the perturbations of the inflaton field. Since the gauge field in our model does not have a vev, demanding that the energy density of the vector field is negligible in the inflationary background is sufficient for this purpose. However, if we assume that inflation has started much earlier than the observed number of e-folds on CMB then the accumulated IR effects of  the vector field perturbations can effectively change the background and put stronger constraints than \eqref{BR-1} and \eqref{BR-2}. This possibility has been studied for example in \cite{Barnaby:2012tk,Fujita:2013qxa} for the case of the kinetic conformal coupling, i.e. $\g=0$. The result of these studies is that the total number of e-folds cannot be much different from the observed number of e-folds on CMB. Here, 
we take the conservative approach that the total number of inflation $\mathcal{N}$
is not very long  in the past. In particular, to solve the flatness and the horizon problems we assume $\mathcal{N} \sim 60$.   As a result, it is enough to only 
consider the backreaction constraints Eqs. \eqref{BR-1} and \eqref{BR-2}.

Besides the above backreaction conditions, we also have to take into account the constraints from the primordial black hole (PBH) formations and the isocurvature perturbations.  
This is somewhat model-dependent and we impose the constraints from PBHs and isocurvature perturbations 
after presenting our models in some details.

\subsection{Relic abundance}

The dynamics after inflation is in principle complicated and needs numerical analysis. One reason is that the oscillating inflaton field can cause production of dark photons during (p)reheating through the couplings $\f$ and $\g$ \cite{Kofman:1997yn,Braden:2010wd}. Specifically, for the hybrid inflationary model, which we have considered in this paper as a working example, the scenario of the reheating is complicated and depends on the chosen values for the couplings $\lambda$ and $g$ in the potential $V(\phi,\psi)$ in Eq.~\eqref{pot} (see Ref.~\cite{GarciaBellido:1997wm} for details).   

To take the problem under theoretical control we assume that the conformal couplings are such that they are stabilized at the end of inflation and afterwards to the trivial values $\f|_{\tau_{\rm e}}=1$ and $\partial_{\tau}\g|_{\tau_{\rm e}}=0$ at least at the background level. With these assumptions the dark photons may not be produced efficiently after inflation and the resultant gauge field energy density at the end of inflation is just redshifted to the late time universe (assuming negligible production/decay during cosmic evolution). In addition, we assume that the reheating takes place instantaneously in which all energy density of the inflaton (minus the increase of the dark photon energy density due to the mass term acquired at the end of inflation) is transferred to the SM fields. 

With these simplifications  the equation for the mode function \eqref{MF-Eq2}  after inflation in terms of cosmic time becomes
\eq{\label{u-after-inflation}
	\ddot{u}_k+H\dot{u}_k+\Big(\frac{k^2}{a^2}+\m^2 \Big)u_k=0\,,
} 
where we have dropped the polarization index in the mode function since the mode function for  both polarizations are the same. 

From Eq.~\eqref{u-after-inflation}, we see that the evolution of the mode function and therefore the evolution of the energy density of the dark photon after inflation depends on the hierarchy among three scales $k/a$, $\m$ and $H$. In the case that the mass scale is subdominant compared to the other two scales,  then the equation is approximately the same as that of a massless vector field and the energy density after inflation evolves as radiation
 $\rho_{\tiny A} \propto a^{-4}$. However, if we assume that the mass scale is dominant then we get the following approximate solution
\eq{
	u_k\propto\frac{1}{\sqrt{a}}\cos(\m t)\,,
}
and then in this regime $\rho_{\tiny A}\propto1/a^3$. For this to be true, the mass scale must be dominant over all relevant momenta, specially the momenta with significant contribution  in the spectrum of the dark photons.  

As mentioned before, $\rho_{\tiny A}$ is related to the total energy accumulated from the gauge field fluctuations at the end of inflation $\rho_{\rm e}$ given in Eq.~\eqref{rhoT5}.  Since  $\rho_{\rm e}$ is constructed from the accumulated energy of all perturbations, we have to check the dominance of the mass scale over all of the modes which exhibit growth during inflation. Let us denote $k_{\rm min}$ and  $k_{\rm max}$ as the first and the last modes which exhibit growth during inflation. Specifically, $k_{\rm min}$ and  $k_{\rm max}$ are the modes that become tachyonic at the initial and end of inflation respectively (see the following section for more details). As we will see $k_{\rm min}$ and  $k_{\rm max}$ are proportional to the scale of the horizon $aH$ at the initial and end of inflation respectively with model dependent coefficients. Then we can roughly write $ k_{\rm min} \sim e^{-{\cal N}} k_{\rm max} $, where ${\cal N}$ is the duration of observed inflation. Therefore, when imposing the conditions above it is actually sufficient to impose the condition on the shortest mode which exhibits growth during inflation, i.e.  $\m \gg k_{\rm max}/a$.  

As a result, we define \emph{non-relativistic} states as those satisfying both conditions $\m\gg H$ and $\m\gg k_{\rm max}/a$ so that $\rho_{\tiny A}\propto a^{-3}$. As a first order approximation, we assume an instantaneous transition from $a^{-4}$ to $a^{-3}$ behaviour at the time $t= t_{\rm NR}$ when both non-relativistic conditions are met. Now the natural question is which one of the conditions $\m \ge H$ and  $\m\ge k_{\rm max}/a$ happens later, guaranteeing the non-relativistic  conditions. Let us parameterize the shortest growing scale as following
\eq{
\label{kappa}
k_{\rm max}\equiv\kappa\, a_{\rm e}H_{\rm e}\,,
}
where $\kappa$ is a dimensionless function of the parameters of the problem which relates $k_{\rm max}$ and the horizon scale at the end of inflation $a_{\rm e}H_{\rm e}$. First of all, note that comparing the two scales we have
\ba
\label{compare}
\frac{k_{\rm max}}{aH} = \kappa \frac{a_ {\rm e} H_ {\rm e}}{a H} \, .
\ea
However, after inflation the universe is decelerating so $a H$ is a decreasing function of time. As a result, if $\kappa>1$ the above ratio is always bigger than unity which means that both non-relativistic conditions are satisfied when the condition $\m\ge k_{\rm max}/a $ is met. However, for parameters that $\kappa<1$ this should be treated with care. More specifically the non-relativistic conditions are satisfied when
\eq{
\frac{a_ {\rm e}}{a_{\rm NR}}=R^{1/4}\,,\quad \text{or} \quad \frac{a_ {\rm e}}{a_{\rm NR}}=\frac{R^{1/2}}{\kappa}\,, 
}  
depending on the fact that for which one $a_{\rm NR}$ is larger, i.e. the condition is satisfied later. Note that for simplicity the nonrelativistic conditions ($\m \gg H$ and $\m\gg k_{\rm max}/a$) are replaced by the simplified conditions $\m = H$ and $\m = k_{\rm max}/a$ at $t= t_{\rm NR}$. While this should suffice for our theoretical treatment here, more accurate numerical solutions must be used for the final relic energy density of dark photons.

We can then imagine different scenarios depending on the values of $R$ and $\kappa$. If $R$ is very large such that $R>\kappa^2>1$, then both conditions are met right at the end of inflation so $t_{\rm NR}$ happens right after reheating. This is a unique feature of our setup based on symmetry breaking which allows for $R>1$ as well as efficient dark photon production during inflation. However, when $R$ is not very large, for example $R<\kappa^2$ or $R<1$, then $t_{\rm NR}$ occurs sometime later than the time of reheating. Note that $t_{\rm NR}$ must not be later than the time of matter-radiation equality for which we must have $\rho_{\tiny A}\propto a^{-3}$ as a genuine dark matter candidate. 
  
As mentioned before, to simplify the analysis we assume an instantaneous reheating for which the temperature $T_{\rm r}$ is given by
\eq{
	\label{He}
	\frac{\pi^2}{30}g_{*,{\rm r}}T_{\rm r}^4=3\Mpl^2H_{\rm e}^2\,,
}
in which $g_{*,{\rm r}}=106.75$ is the number of relativistic degrees of freedom for energy density at the time of reheating. The energy density of the dark photon evolves like $a^{-4}$ before the time $t_{\rm NR}$ and evolves like $a^{-3}$ after that. As a result the energy density of the dark photon today $\rho_{\tiny A,0}$ will be
\eq{
\rho_{\tiny A,0}=\rho_{\rm e}\Big(\frac{a_ {\rm e}}{a_{\rm NR}}\Big)\Big(\frac{g_{s*,0}}{g_{s*,{\rm r}}}\Big)
\Big(\frac{T_0}{T_{\rm r}}\Big)^3\,,
} 
where $T_0\sim10^{-13}$GeV is the CMB temperature today, $T_{\rm r}$ is the reheating temperature, $g_{s*,0}=3.9$, $g_{s*,{\rm r}}=106.75$ are the relativistic degrees of freedom for entropy density today and at the time of reheating. Finally, $\rho_{\rm e}$ is the energy density of dark photons at the end of inflation given by Eq.~\eqref{rhoT5}. 
Plugging the above values,  the relic dimensionless energy density today of dark photon  then is given by
\ba 
\label{relic-DM}
\Omega_{\tiny A,0} \equiv \frac{\rho_{\tiny A,0}}{3\Mpl^2 H_0^2}&=& \Big(\frac{g_{s*,0}}{g_{s*,{\rm r}}}\Big)\Big(\frac{\pi^2}{90}g_{*,{\rm r}}\Big)^2\Big(\frac{T_0^4}{3\Mpl^2 H_0^2}\Big) \, 
\Big(\frac{a_ {\rm e}}{a_{\rm NR}}\Big)\Big(\frac{T_{\rm r}^5}{\Mpl^4T_0}\Big)\mathcal{C} \nonumber\\
&=&
\label{relic-DMfinal}
 0.5\times10^{-4}\Big(\frac{a_ {\rm e}}{a_{\rm NR}}\Big)\Big(\frac{T_{\rm r}}{10^{12}{\rm GeV}}\Big)^5\mathcal{C}\,,
\ea
where the parameterization Eq. (\ref{rho-param}) has been used and 
we have used Eq.~\eqref{He} to write $H_{\rm e}$ in terms of reheating temperature. In each model of inflation with a vector field one can compute $\mathcal{C}$, $\kappa$, $R$ and $T_{\rm r}$. Then the general expression \eqref{relic-DMfinal} gives the relic energy density of dark photons. It is useful to have an expression for the mass of the dark photon in terms of the reheating temperature and the ratio $R$. Assuming instantaneous reheating from Eq.~\eqref{He} we have
\eq{
\label{mAA}
\m=3.4\times10^6\sqrt{R}\Big(\frac{T_{\rm r}}{10^{12}{\rm GeV}}\Big)^2\,{\rm GeV}\,.
}  
An important constraint is that non-relativistic conditions must be met before the time of matter-radiation equality. This imposes a lower lower bound on $R$ as following
\eq{
\label{R-cond1}
R\gtrsim\max\Bigg\{\frac{10^{-43}}{\kappa^2},10^{-86}\Big(\frac{10^{12}\,{\mbox{GeV}}}{T_{\rm r}}\Big)^2\Bigg\}\times\Big(\frac{10^{12}\,{\mbox{GeV}}}{T_{\rm r}}\Big)^2\,,
}  
where we have used $T_{\rm eq}\sim10^{-9}$GeV as the matter-radiation equality temperature. As an estimation of lower bound on $R$, suppose we take the minimum reheat temperature allowed from Big Bang Nucleosynthesis, $T_{\rm r} \sim 10$MeV and set $\kappa\sim 1$. Then the above expression requires $R\gtrsim 10^{-15}$. From Eq.~\eqref{mAA}, this corresponds to a lower bound of $\m\sim10^{-20}$eV. Of course, by taking higher value of the reheat temperature, say $T_{\rm r} \sim 10^{12} \mbox{GeV}$, one can take lower values for $R$. However, masses lower than $\sim10^{-24}$eV are ruled out due to the existence of dark matter dominated structures of length scale $\sim 1$kpc \cite{Nelson:2011sf,Kolb:2020fwh}.

Before closing this section let us examine the constraint on the model parameters 
from the CMB upper bound on the tension of cosmic strings.  As mentioned before, the production of cosmic strings  is a natural outcome of our setup with the tension $\mu$ given in Eq. (\ref{string-mu}).  
Using the definitions of $m_A^2$ and $R$ in Eqs. (\ref{mA})  and (\ref{R}) and the formula for the reheat temperature Eq. (\ref{He}) we can obtain a relation between $\mu$, $R$ and $T_{\rm r}$ as follows 
\ba
R \Big(\frac{T_{\rm r}}{M_P} \Big)^4 \sim \frac{720\,  \e^2}{\pi g_{*, {\rm r}}} G \mu 
\sim  2  \e^2 G \mu \, .
\ea
Using the upper bound $G \mu \lesssim 10^{-7}$ from the CMB observations,  
the above expression yields the following upper bound 
\ba
\label{R-bound}
R \Big(\frac{T_{\rm r}}{M_P} \Big)^4 \lesssim 10^{-7} \, .
\ea
In the final estimation we also used the assumption that $\e^2 <1$. This is because $\e$ is the effective gauge coupling at the end of inflation and in order to have the perturbative QFT of gauge field under control we require $\e<1$. 

We see that the bound Eq. (\ref{R-bound}) allows for a wide rang of the value of $R$. For example, taking $T_{\rm r} = 10^{12}\,  {\rm{GeV}}$ we only require $R < 10^{17}$ to satisfy the observational  bound on the tension of strings. Increasing the reheat temperature the allowed values of $R$ become smaller. Taking the extreme upper bound  $T_{\rm r}= 10^{15} \, {\rm{GeV}}$, we require $R <10^5$.  On the other hand, by reducing $T_{\rm r}$ the upper bound on $R$ becomes much more relaxed. For example, taking the lower bound $T_{\rm r} \sim 10$MeV we only require 
$R < 10^{73}$ to satisfy the bound on the cosmic string tension. However, note that in terms of the mass of the dark photon, the condition \eqref{R-bound} corresponds to an upper bound $\m \lesssim 10^{-3}\Mpl\sim10^{15}$GeV. 

The above considerations show that our setup supports dark photons with a wide range of masses from $10^{-20}$eV up to $10^{15}$GeV. While light dark matter candidates are quite popular, superheavy dark matter are not usually considered. In fact there are a couple of studies which consider superheavy dark matter particles \cite{Chung:1998zb,Chung:1998bt,Kannike:2016jfs,Babichev:2018mtd,Li:2020xwr,Kolb:2020fwh,Allahverdi:2020uax}. One reason is the unitarity bound on dark matter particle mass $\lesssim10^5$GeV \cite{Griest:1989wd}. However, this is limited to dark matter models that are in thermal equilibrium with the cosmic plasma and are produced via freeze-out mechanism. There is no such bound on non-thermal scenarios like particle production we consider here. In passing, it is worth to mention that the typical energy of the particles produced thermally is $E\sim T$ where $T$ is the temperature of the plasma at the time of production \cite{Kolb:2020fwh}. As a result, light dark photons in our scenario can become hot (relativistic) if they can interact efficiently with the thermal bath. However, this should not be an issue for heavy dark photons. 

The main concern for superheavy dark photons is that they can potentially decay to other particles after inflation. In fact any viable dark matter model must be cosmologically stable. The most studied interaction for the dark photon is the kinetic mixing term
\eq{
\mathcal{L}\supset-\frac{1}{2}\varepsilon F_{\mu\nu}F^{\mu\nu}_{\rm \scriptscriptstyle{EM}}\,,
}    
which is a dimension 4 and renormalizable operator and which is allowed by symmetry.  In the above interaction, $F^{\mu\nu}_{\rm \scriptscriptstyle{EM}}$ is the gauge strength tensor for the SM photons and $\varepsilon$ is a dimensionless coupling (not to be confused with the polarization vectors). Equivalently, one can choose to diagonalize the kinetic part of the action by redefining the two vector fields but the price is that the SM electromagnetic current also becomes charged under the dark photon with the scaled charge $-\varepsilon\e_{\rm \scriptscriptstyle{EM}}$ \cite{Fabbrichesi:2020wbt,Nelson:2011sf}. The generation  of dark photons through the kinetic mixing interaction has been under vast experimental studies and there are strong upper bounds on the numerical value of $\varepsilon$ for different range of masses \cite{Essig:2013lka,Fabbrichesi:2020wbt}. The lack of detection of dark photons so far implies that the coupling $\varepsilon$, if nonzero, must be extremely tiny. Note that this is not natural because even if the kinetic mixing is set to zero at the tree level, it will usually be generated at the 1-loop order with $\varepsilon\sim10^{-2}$ \cite{Holdom:1985ag}. One can imagine many scenarios to make the natural value of the kinetic mixing term tiny to reconcile it with the experiments. These include modifying the UV physics such as considering suitable string compactification or larger GUT gauge group, or demanding further suppression in loop calculations via multi-loop mechanisms \cite{Gherghetta:2019coi}.

\fg{
	\centering
	\includegraphics[width=0.3\textwidth]{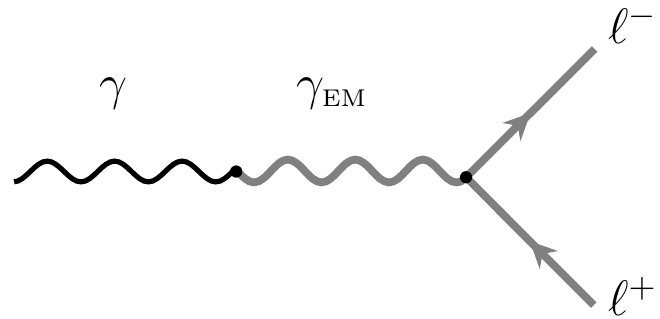}
	\hspace{8mm}
	\includegraphics[width=0.255\textwidth]{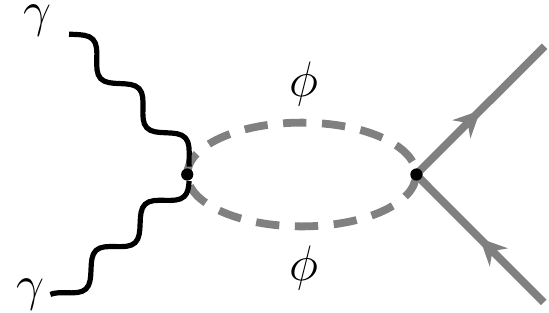}
	\hspace{8mm}
	\includegraphics[width=0.18\textwidth]{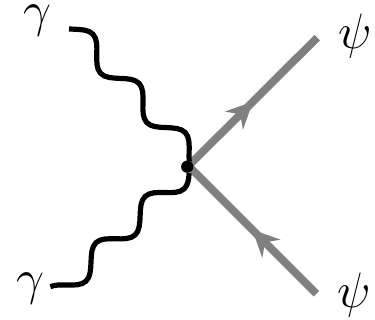}
	\caption{Possible decay channels for the dark photon dark matter. Decay to the SM leptons through the kinetic mixing (left). Kinetic coupling to inflaton induces a channel of decay to SM particles (center). Dark photon can decay to the waterfall field if it is heavier than the waterfall (right).}
	\label{fig:feynman}
}

The left diagram of Fig.~\ref{fig:feynman} is the leading decay channel of the dark photon to the SM charged particles through the kinetic mixing interaction. We demand that the decay rate is smaller than the present value of the Hubble parameter such that the dark photon has a life-time greater than the age of the universe. Assuming the dark photon is extremely heavy results in the decay rate
\eq{
\label{eps}
\Gamma\approx\frac{1}{3}\alpha_{\rm \scriptscriptstyle{EM}}\varepsilon^2\m\lesssim H_0\qquad\implies\qquad \varepsilon\lesssim10^{-22}\left(\frac{10^5{\rm GeV}}{\m}\right)^{1/2}\,,
}
where $\alpha_{\rm \scriptscriptstyle{EM}}=\frac{1}{137}$ is the structure constant. As is clear in the above derivation, heavier dark photons require even smaller values of $\varepsilon$. Note that the above decay channel is only applicable for dark photons heavier than at least two electrons ($\sim$MeV). For lighter masses, the above bound is relaxed by many orders of magnitude as the most efficient decay channel would be into three SM photons through a fermion loop with extremely low rate. Further, comparison of the life-time with the age of the universe gives but only a lower bound. The possibility of dark matter decay into SM particles usually puts stronger bounds on the life-time through cosmological and astrophysical observations \cite{Zhang:2007zzh,PalomaresRuiz:2007ry,Cembranos:2007fj}. This can decrease the upper bound of the kinetic mixing parameter in Eq.~\eqref{eps} a few orders of magnitude.  

Although in minimal models such tiny values are not favored, it is possible to consider UV completion setups that may  result in very small $\varepsilon$. In fact, there are not much experimental bounds on $\varepsilon$ for superheavy dark photons. However, bounds on lighter masses are quite below the natural value which motivates considering tiny $\varepsilon$. Therefore, we keep the wide mass range $\m \in [10^{-20}$eV$-10^{15}$GeV$]$ in this paper. Once we have a stronger bound on $\varepsilon$ for the heavy dark photons, we only need to consider those regions for the mass of the dark photon which are consistent with that bound.

Other possible decay channels might result from dark photon coupling to the inflaton and the waterfall fields. Remember that we needed the kinetic conformal couplings $\f(\phi)$ and $\g(\phi)$ in order to break the conformal invariance of the dark photon sector to sufficiently produce them during inflation. We have assumed that at the background level they settle to their final values $\f=1$ and $\g={\rm const.}$ after inflation. However, at the perturbative level they can be the sources for the dark photon decay to SM particles through the coupling of inflaton to SM. The latter is needed to reheat the universe after inflation. This channel is depicted in the center diagram in Fig.~\ref{fig:feynman} \footnote{For the diagram in Fig.~\ref{fig:feynman} we have chosen $\f(\phi)\sim\phi^2$ but complicated functions yield  more complicated diagrams.}. Another possibility is the decay into the waterfall particles. Remember that the waterfall field was charged under the dark photon with charge $\e$. After the symmetry breaking takes place, the interaction between waterfall particles around the new vacuum and the dark photon is of the form $\e^2\psi^2A_\mu A^\mu$. This vertex is shown in the right diagram of Fig.~\ref{fig:feynman}. 

The decay to the inflaton or the waterfall fields happens if the dark photon is heavier than them. After the symmetry breaking the masses of the dark photon, inflaton and the waterfall are
\eq{
\m=\frac{\e}{\sqrt{\lambda}}M\,,\qquad m_\phi=\sqrt{m^2+\frac{g^2M^2}{\lambda}}\approx\frac{g}{\sqrt{\lambda}}M\,,\qquad m_\psi=\sqrt{2}M\,. 
}  
As a result, it is clear that if we set $\e\lesssim g$ then the decay into inflaton cannot take place. Otherwise, one has to compute the decay rate which depends on the details of the reheating theory as well and we do not pursue this in this paper anymore. Similarly, if one sets $\e\lesssim\sqrt{2\lambda}$ then the decay into the waterfall field does not occur as well. Note that even in this conservative regime, the parameter $R$ defined in Eq.~\eqref{R} can still be large due to the largeness of the parameter $\beta$. Finally, we note that all possible decay channels through the graviton loops are suppressed with the Planck scale as all matter fields are minimally coupled to gravity in our model. In the rest of the paper we assume that all decay channels have negligible effect on the evolution of the dark photon energy density.

Before closing this section we provide a pictorial  summary of our model in Fig.~\ref{fig:diagram}. In this figure, we have shown the evolution of the energy density of different fields in the model \footnote{We thank the anonymous referee who suggested to add this graph for a better clarification on the model.}. The period of inflation is driven by the vacuum energy of the waterfall field $\psi$. However, at the end of inflation it can transfer most of its energy to the inflaton field depending on the parameters of the inflationary model (see Ref.~\cite{GarciaBellido:1997wm} for details). As a result, we have plotted the total energy density of the inflaton$+$waterfall system. This is mostly transferred to the radiation during reheating period. The radiation is redshifted as $a^{-4}$ by cosmic expansion. During inflation the transverse modes of the vector fields get excited due to non-adiabatic particle production in the inflationary background. At the end of inflation there is a sharp increase in  its energy   due to the Higgs mechanism. After inflation, the transverse modes evolve like radiation until $t_{\rm NR}$ and like matter after that, assuming negligible other forms of decay. Finally, note that the longitudinal mode is not present during inflation and is not excited after that. 

In the following sections we will consider specific models within the general setup considered above. We first compute $\mathcal{ C}$ and $\kappa$ for each model. Then we need to look at the constraints from the backreaction conditions \eqref{BR-1} and \eqref{BR-2} as well as other possible constraints coming from the dynamics of perturbations to know the legitimate range of parameters. Finally we will compute the present energy density of dark photons. 

\fg{
	\centering
	\includegraphics[width=0.67\textwidth]{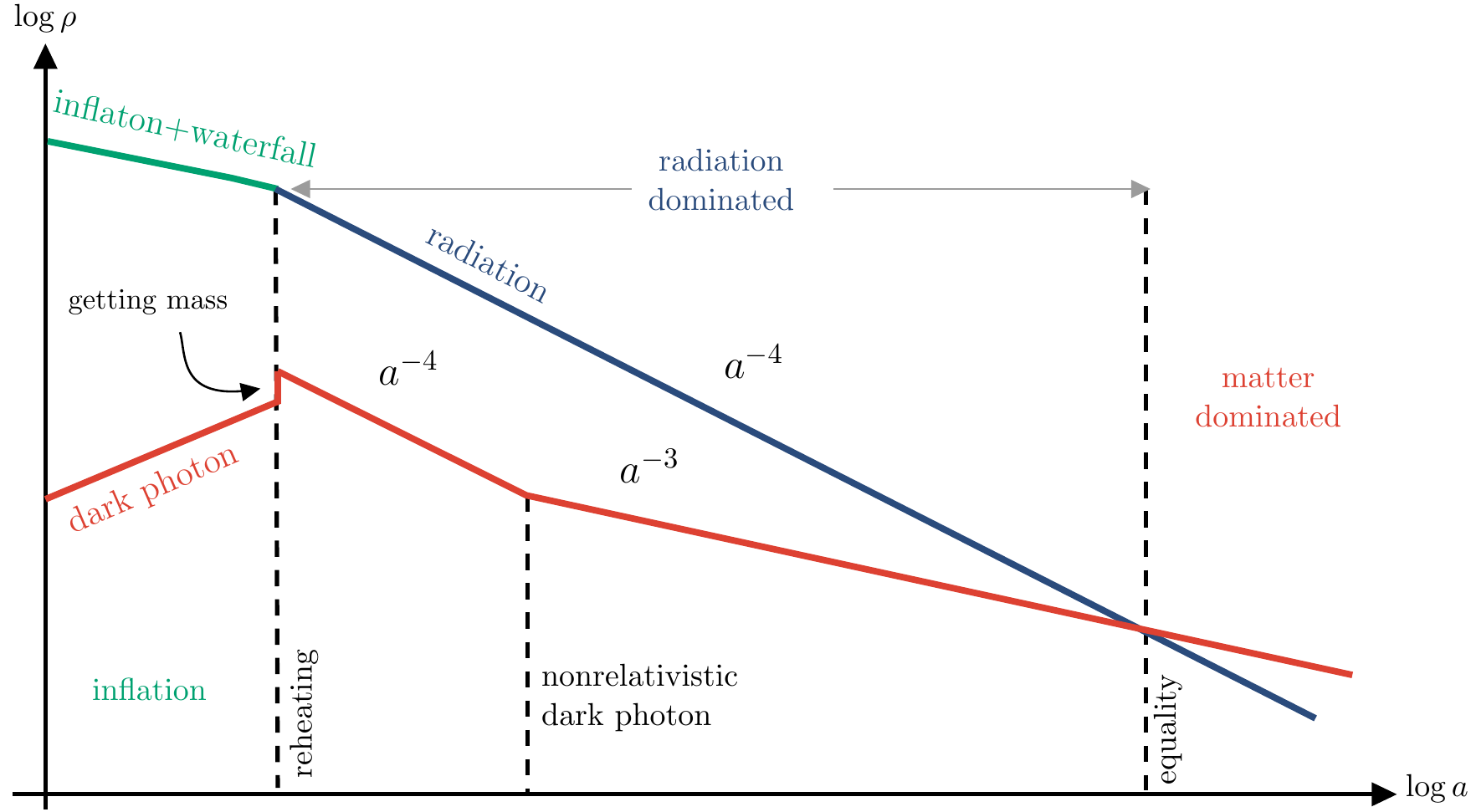}
	\caption{Time evolution of different fields in our rather general inflationary model.}
	\label{fig:diagram}
}

\section{The specific model}\label{sec-Model}

As we mentioned before, during inflation the gauge field energy density decays quickly and this issue can be remedied if energy is pumped from the inflaton field to the gauge field through the couplings $\f(\phi)$ and $\g(\phi)$ in the action \eqref{action}. Naturally, the conformal couplings $f$ and $h$ are functions of the inflaton field $\phi$. However, during inflation the scale factor is related to the evolution of $\phi$ as 
\begin{equation}\label{a}
a(\phi) \approx \exp\bigg[-\frac{1}{\Mpl^2
}\int\frac{V}{V_{,\phi}}d\phi\bigg] \,,
\end{equation}
in which $V$ is given by Eq.~\eqref{V}. Therefore, we can imagine that the
 conformal couplings will be functions of $a$ as well. Motivated from the past works on  models of anisotropic inflation and the mechanism of magnetogenesis during inflation, we consider the following phenomenological ansatz for the conformal couplings 
\begin{equation}\label{fg-a}
\f(\phi)=\f_{\rm e} \left(\frac{a}{a_{\rm e}} \right)^{-p} \,, \hspace{1cm} 
\g(\phi)=\g_{\rm e} \left(\frac{a}{a_{\rm e}} \right)^{-q} \,,
\end{equation}
in which  $\f_{\rm e}$ and $h_{\rm e}$ are the final values of the conformal couplings at the end of inflation.  Alternatively, using the relation $a\simeq -(H\tau)^{-1}$ during inflation we can express the conformal couplings in terms of the conformal time as follows 
\begin{eqnarray}\label{f1}
\f(\tau) = \f_{\rm e} \left(\dfrac{\tau}{\tau_{\rm e}}\right)^{p}\,, \hspace{1cm} 
\g(\tau) = \gamma \f_{\rm e}\left(\dfrac{\tau}{\tau_{\rm e}}\right)^{q} \,,
\end{eqnarray}
where we have parametrized $\g_{\rm e}=\gamma\f_{\rm e}$ by means of a constant parameter $\gamma$ without loss of generality. 

As mentioned before,  the couplings of the form Eq. (\ref{f1}) are common in the context of magnetogenesis \cite{Ratra:1991bn,Garretson:1992vt,Anber:2006xt, Martin:2007ue, Demozzi:2009fu,Kanno:2009ei,Emami:2009vd,Fujita:2012rb,Caprini:2014mja,Caprini:2017vnn, Schober:2020ogz, Talebian:2020drj}, models of anisotropic inflation  \cite{Watanabe:2009ct, Watanabe:2010fh, Bartolo:2012sd, Emami:2010rm, Emami:2013bk, Emami:2015qjl}, inflation  based on a pseudoscalar \cite{Anber:2006xt,Sorbo:2011rz}, multiple vector field models \cite{Yamamoto:2012tq,Yamamoto:2012sq,Firouzjahi:2018wlp,Gorji:2020vnh} and Schwinger mechanism during inflation \cite{Kobayashi:2014zza,Lozanov:2018kpk,Shakeri:2019mnt}.  In particular, the so-called anisotropic inflation corresponds to $p \simeq 2$ and $\gamma=0$ but with a nonzero background value for the gauge field \cite{Watanabe:2009ct}. Moreover, as we will show, the case with $p=0$ and $q\ll1$ corresponds to  inflation  based on a pseudoscalar field \cite{Sorbo:2011rz}. Of course, we can consider the case with $p\neq0$ and $\gamma\neq0$ and look for the effects of both terms simultaneously which is the subject of subsection \ref{sec-model-general}. 

As inflation ends, $\f$ and $\g$ reach their final values $\f_{\rm e}$ and $\g_{\rm e}$, and stop evolving. We scale $\f_{\rm e}$ to be unity which leads to $\g_{\rm e}=\gamma$. The kinetic term for the gauge field then takes the standard form and the parity violating term becomes a total derivative term so after inflation the dynamics of the gauge field is given by the Maxwell theory. An important criteria to take into account is that the effective gauge coupling is $\f^{-1}$. In order to have a perturbative theory, we require $\f^{-1} <1$ throughout inflation, otherwise the theory becomes strongly interacting and the perturbative analysis cannot be trusted. This criteria prohibits negative values of $p$. So, in order to bypass the strong coupling problem we assume $p>0$ throughout our analysis. Note that there is not such a constraint on the value of $q$.

Assuming the specific form of the couplings given in Eq. (\ref{f1}), the mode function Eq.~\eqref{MF-Eq} takes the following form
\ba
\label{v-eq1}
v_{k,\lambda}'' + \Big( k^2 + 2 \lambda k \frac{\xi}{\tau} -\frac{p (p-1)}{\tau^2} \Big) v_{k,\lambda}=0\,; 
\hspace{1cm} \xi \equiv - \gamma^2 q\Big(\dfrac{\tau}{\tau_{\rm e}}\Big)^{2(q-p)}\approx\text{constant} \,,
\ea
where we have assumed that $|q-p|\ll1$ so that we can ignore the time dependence of the parameter $\xi$ and treat it as a constant. As before, the special case of kinetic conformal coupling corresponds to $\xi=0$ for which in the particular case of $p=2$ we find the usual term $2/\tau^2$ in the mode function frequency which is the hallmark of the massless scale invariant perturbations in a fixed de Sitter background. In the case of $p=0$ we have the axion-like coupling of the inflaton to the vector field similar to the setup of pseudoscalar inflation \cite{Sorbo:2011rz} where the time dependence of $\xi$ is negligible to first order of slow-roll parameter. Note that such a model is considered before for example in Refs. \cite{Caprini:2014mja,Caprini:2017vnn}, where they have set $q=p$ and arrived at a similar equation of motion as Eq.~\eqref{v-eq1}. However, in that case the limit of $p\to0$ becomes nontrivial while with our definition we can obtain all limits consistently.    

The solution for the mode function can be obtained from Eq.~\eqref{v-eq1} assuming the Bunch-Davies initial condition as following
\ba
\label{mode}
v_{k,\lambda} (\tau) = \frac{e^{\lambda\pi\xi/2}}{\sqrt{2k}} W_{\mu,\nu}\big(2ik \tau\big) \,; \hspace{1cm}
\mu\equiv-i\lambda\xi \,, \hspace{.5cm} \nu\equiv p-1/2\,,
\ea
in which $W_{\mu,\nu}(z)$ is the Whittaker function. By using the asymptotic behaviour $W_{\mu,\nu}(z)\to z^\mu e^{-z/2}$ for $z\to\infty$, one can show that $v_{k,\lambda} (\tau)$ approach the standard Bunch-Davies solutions at early times. Further,  $W_{\mu,\nu}(z)\to z^{1/2-\nu} \Gamma(2\nu)/\Gamma(\nu+1/2-\mu)$ for $z\to0$ so the 
 super-horizon behaviour is given by 
\eq{
v_{k,\lambda}\to \frac{e^{\lambda\pi\xi/2}}{\sqrt{2k}}e^{i\pi(2\nu-1)/4}\frac{\Gamma(2p-1)}{\Gamma(p+i\lambda\xi)}(-2k\tau)^{1-p}\,,\qquad -k\tau\to0\,.
}   

First of all, we see the dependence on $\xi$ is always through the combination $\lambda\xi$ and, therefore, we assume that $\xi>0$ without loss of generality. As we see, a nonzero value of $\xi$ induces chirality in the spectrum of the produced dark photons so that one of the polarizations is enhanced significantly compared to the other. Furthermore, from the point of view of the equation of motion (and also from the properties of the Whittaker function) we see that the transformation $p\to1-p$ (or equivalently $\nu\to-\nu$) does not change the properties of the mode function. Since we are interested in $p\geq0$ to avoid the strong coupling limit, it is then sufficient to study only the solution for $p\geq1/2$ (or equivalently $\nu\geq0$) as the region $0\leq p\leq1/2$ is the same as $1/2\leq p\leq1$. Note also that the limit of axion-like coupling, i.e. $p=0$ is the same as $p=1$ (or equivalently $\nu=1/2$). 
However, we are ultimately interested in the energy density which unlike the mode function explicitly depends on $p$ through the conformal factor $\f(\phi)$ (see for example \eqref{rhoT3}). As a result, we need to consider the whole region $p\geq0$ and the true limit of axion-like coupling is $p=0$.      

We are interested in the regime where the mode function grows until the end of inflation. This is most efficient if the mode function undergoes tachyonic growth. To see this more clearly, we define a new variable $x\equiv-k\tau$ in term of which Eq.~\eqref{v-eq1} takes the form
\eq{\label{Eq-tachyonic}
\dv[2]{v_{k,\lambda}}{x} + \Big( 1 - 2 \frac{\lambda\xi}{x} -\frac{p (p-1)}{x^2} \Big) v_{k,\lambda}=0\,.
}   

We see that depending on the values of the parameters $p$ and $\xi$, we may have completely different scenarios. If $\xi^2+p(p-1)<0$, which is possible only for $0\leq p<1$, the frequencies for both polarizations are always positive and therefore there is no tachyonic regime. Note that this is different than the case $\xi=0$ where we also do not have any tachyonic growth. On the other hand, tachyonic growth occurs if $\xi^2+p(p-1)\geq0$ which is possible for all $p>0$. In this case, we have two different regimes during which tachyonic growth happens. First, we note that the condition $\xi^2+p(p-1)\geq0$ is satisfied for $p\geq1$ independent of the value of $\xi$. In this regime, which we call \emph{kinetic-like} regime, the momenta which start becoming tachyonic right from the beginning of inflation at $\tau_{\rm i}$ until the end of inflation at $\tau_{\rm e}$ are given by\footnote{\label{footnote:diffusion}We assume that $k_{\rm max}$ is lower than the comoving wave number corresponding to the diffusion length of the dark photon, which is unknown and depends on the dark electric conductivity and thus on the nature of the interactions of the dark photon with other fields in the dark sector.}
\eq{
	\label{ks}
	k_{\rm min}=\frac{x_\lambda}{-\tau_{\rm i}}\,,\hspace{1cm} k_{\rm max}
	=\frac{x_\lambda}{-\tau_{\rm e}}\,,\qquad(\text{kinetic-like}, \,\, p\geq1)\,,
}  
where we have defined $x_\lambda\equiv\lambda\xi+\sqrt{\xi^2+p(p-1)}$. Note that $x_\lambda$ depend on the polarization for nonzero $\xi$. We will see that the energy density of the dark photon will be chiral in this regime and the chirality is controlled by $\xi$. Thus, for small $\xi$ both polarizations play roles. The special case of $\xi=0$ corresponds to the standard kinetic conformal coupling (as in anisotropic inflation)  which is the reason why we called this regime kinetic-like. We will consider this limit in details in the following section. Note that in the kinetic-like case $\kappa=x_\lambda$ where $\kappa$ is defined in Eq.~\eqref{kappa}.

The second regime is defined by the condition $\xi^2+p(p-1)\geq0$ and $0\leq p<1$. In this regime, the parameter $\xi$ is restricted as $\xi^2\geq p(1-p)$ and only the plus polarization becomes tachyonic. We thus reasonably call this regime \emph{chiral-like}. The range of momenta that experience tachyonic growth in this regime are given by (see footnote \ref{footnote:diffusion})
\eq{
	\label{ks-t}
	k_{\rm min}=\frac{{\tilde x}_-}{-\tau_{\rm i}}\,,\hspace{1cm} k_{\rm max}
	=\frac{{\tilde x}_+}{-\tau_{\rm e}}\,,\qquad(\text{chiral-like},\,\, 0\leq p<1)\,,
}
where we have defined ${\tilde x}_\pm=\xi\pm\sqrt{\xi^2-p(1-p)}$. Note that up to here everything depends only on the properties of the equation of motion and therefore the results for the regions $0\leq p\leq1/2$ and $1/2\leq p\leq1$ would be similar. In this case $\kappa={\tilde x}_+$. 

We are interested in the energy density at the end of inflation given in Eq.~\eqref{rhoT5}, which following the parameterization \eqref{rho-param} and after using Eq.~\eqref{mode}, turns out to be
\eq{
\spl{
\label{F}
\mathcal{C}(p,\xi,\mathcal{N}) = \frac{1}{8\pi^2} \sum_\lambda e^{\lambda\pi\xi}
\int_{x_{\rm min}}^{x_{\rm max}} &\dd{x} x \bigg\{ \abs{W_{\mu+1,\nu}(2ix)}^2+\Big(\frac{R}{f^2}+p^2+\xi^2-2\lambda\xi x+2x^2 \Big)\abs{W_{\mu,\nu}(2ix)}^2 \\
&+2{\rm Re}\Big[\big( p+ix-i\lambda\xi \big)
W_{\mu,\nu}(-2ix) W_{1-\mu,\nu}(2ix)\Big]\bigg\}\,.
}
}
Here, the range of integration is 
\eq{
	\label{intreg1}
	x_{\rm min}=x_\lambda e^{-\mathcal{N}}\,,\hspace{1cm} x_{\rm max}=x_\lambda\,,\qquad(\text{kinetic-like},\,\, p\geq1)
}  
for the kinetic-like regime and 
\eq{
	x_{\rm min}={\tilde x}_-e^{-\mathcal{N}}\,,\hspace{1cm} 
	x_{\rm max}={\tilde x}_+\,,\qquad(\text{chiral-like},\,\, 0\leq p<1)
}
for the chiral-like regime. Note that in the former case we have summed over both polarizations  as both polarizations are excited  while in the latter case we only kept the dominant one which is $\lambda=+$ in our convention. 

It can be shown that for $p>2$ the dominant contribution of the integral (\ref{F})
comes from its lower limit which is very small. As a result, it is sufficient to consider the leading contribution of the integrand in the limit of $x\to0$. In this case, Eq.~\eqref{F} simplifies to 
\begin{equation}
\label{appp2}
\mathcal{C}(p,\xi,\mathcal{N}) \approx \frac{\Gamma (2 p - 1 )^2}{4^{p+1}\pi^2} 
\Big[(2 p - 1 )^2+\frac{R}{f^2}\Big] \Big(\sum_\lambda 
\frac{e^{  \lambda \pi  \xi }}{x_\lambda^{2p-4}|\Gamma \left(p - i \lambda  \xi \right)|^2}\Big)\frac{e^{(2p-4)\mathcal{N}}}{p-2} \,,\qquad(p>2)\,,
\end{equation}
which shows strong dependence on the number of e-folds $\mathcal{N}$. However, for $p\leq2$ we need to compute the integral numerically. 

Before moving on to compute the relic energy density first we need to constrain the parameter space of the theory by demanding negligible backreactions on the background dynamics. The first backreaction constraint Eq. \eqref{BR-1} can be written in the form  
\eq{
\label{BR11}
\frac{8\pi^2}{3}\epsilon\mathcal{P}\mathcal{ C} {|_{R=0} }\ll1\,,
}
where  we have used the definition of the power spectrum of curvature perturbations
\eq{
	\mathcal{P}=\frac{H^2}{8\pi^2\epsilon\Mpl^2}\approx2\times10^{-9}\,,
}
and $\epsilon=\dot{\phi}^2/2\Mpl^2H^2$  is the slow-roll parameter. Note that since in the backreaction condition Eq. \eqref{BR-1}  we use $\rho_T$ and 
not $\rho_{\rm e}$ we have to use the value of $\mathcal{ C}$ during inflation when the gauge field is still massless, corresponding to $R=0$. 

The second backreaction parameter comes from the equation of motion of the inflaton field as discussed in Eq.~\eqref{BR-2}. If we parametrize $S_\phi\equiv(H/\dot{\phi})H^4\Sigma$ then the second condition can be written as
\eq{
\label{BR22}
\frac{4\pi^2}{3}\mathcal{P}\Sigma\ll1\,.
}
By using Eqs.~\eqref{SS} and \eqref{mode} we find
\begin{eqnarray}\label{G}
&&\Sigma(p,\xi,\mathcal{N}) = \frac{1}{4\pi^2} \sum_\lambda e^{\lambda\pi\xi}
\int_{x_{\rm min}}^{x_{\rm max}}\dd{x} x 
\bigg\{ p\abs{W_{\mu+1,\nu}(2ix)}^2+p\Big(p^2+\xi^2-4\lambda\xi x \Big)\abs{W_{\mu,\nu}(2ix)}^2
\nonumber \\ && \hspace{4cm} 
+2{\rm Re}\Big[\big( p(p+ix)-i\lambda\xi(p-ix) \big) W_{\mu,\nu}(-2ix) W_{1-\mu,\nu}(2ix)\Big]\bigg\}. 
\hspace{.7cm}
\end{eqnarray}
In the case that we have $\mathcal{C}\sim \Sigma$, the condition \eqref{BR22} is more restrictive than \eqref{BR11} by a factor of $\epsilon$. Assuming that this is the case we have plotted the logarithm of the left hand side of Eq.~\eqref{BR22} for different values of $p$ and $\xi$ in Fig.~\ref{fig:back1}.  
\fg{
	\centering
	\includegraphics[width=0.5\textwidth]{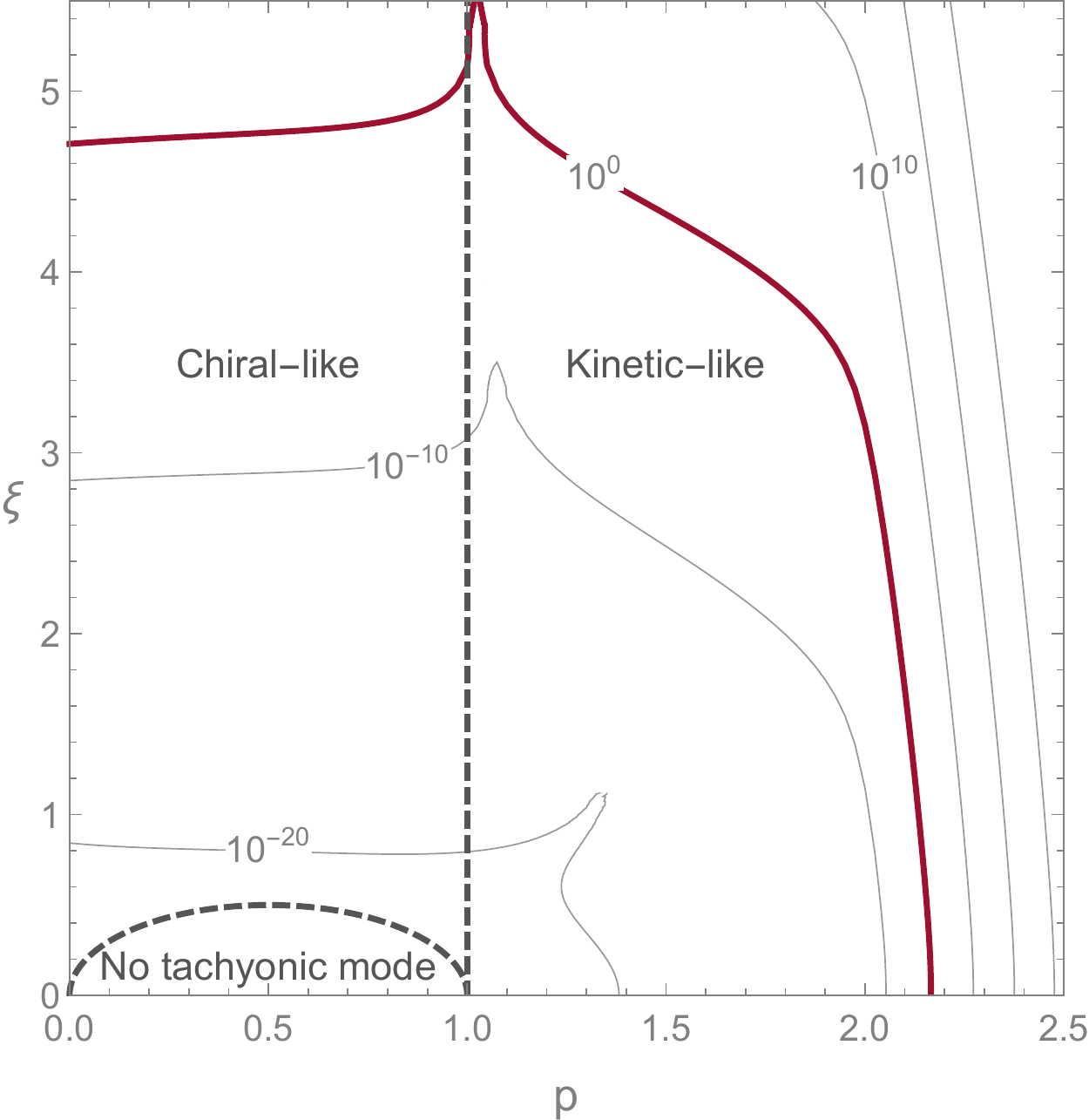}
	\caption{The contour plot of the backreaction constraint  Eq.~\eqref{BR22}  in terms of  $p$ and $\xi$.  Each curve corresponds to a fixed  value of the left hand side of Eq.~\eqref{BR22}. The thick red curve shows the maximum limit at hand. We can see that for all interesting values we must have $p\lesssim2.2$ and $\xi\lesssim4.6$ to avoid backreaction. We have set the number of e-folds $\mathcal{N}=50$.}
	\label{fig:back1}
}
This shows that we need to have $p\lesssim2.2$ and $\xi\lesssim4.6$ to respect the backreaction conditions. The former can be understood from the approximate result of Eq.~\eqref{appp2} for $p>2$ which has exponential dependence on ${\cal N}$ for 
$p>2$. 

Besides the above backreaction conditions we also need to make sure that the isocurvature perturbations do not violate the observational constraints. From our 
expression for $\rho_{\rm e}$ in \eqref{rhoT3} and the parameterization in \eqref{rhoT5}, the isocurvature  perturbation can be estimated as
\eq{
\label{isoc}
s= \frac{\delta {\cal C}(p, R, {\cal N} )}{ {\cal C}} \simeq
\frac{1}{ {\cal C}}\pdv{{\cal C}}{\mathcal{N}}\delta\mathcal{N}=\frac{1}{{\cal C}}\pdv{{\cal C}}{\mathcal{N}} {\cal R}\,,
}
where we have related the perturbation in the number of e-folds to the curvature perturbation ${\cal R}$ through the so-called $\delta{\cal N}$ formalism \cite{Sasaki:1995aw, Lyth:2004gb, Wands:2000dp, Abolhasani:2019cqw}. Note that we also have the contribution from the variation 
of $R$. However, the variations  in $R$ is related to the fluctuations of the waterfall field which are heavy and are suppressed, so the contribution from $\delta R$ in the above isocurvature perturbations is suppressed.   

Interestingly,  Eq. (\ref{isoc}) indicates that the isocurvature and the curvature perturbations are fully correlated. 
As a result, the ratio of the dark matter isocurvature perturbations power spectrum to the curvature perturbations power spectrum is obtained to be
\eq{
\beta\equiv\frac{\mathcal{P}_s}{\mathcal{P}}=\left(\frac{1}{{\cal C}}\pdv{{\cal C}}{\mathcal{N}}\right)^2  \lesssim10^{-3}\,,
} 
where we have used the recent upper bound from Planck data for the fully correlated dark matter isocurvature perturbations  \cite{Akrami:2018odb}. The strong dependence on the number of e-folds only occurs for $p>2$  for which from Eq.~\eqref{appp2} we obtain  $\beta = (2 p-4)^2$ yielding  the  upper bound
\eq{\label{bound-p2}
p\lesssim2.01\,.
}
As a result, the constraint on $p$ from the isocurvature perturbations is stronger than what we already have obtained from the backreaction condition. Such a strong constraint on $p$ motivates us to focus on the values $p\leq2$ in the rest of the paper. Note that the main reason, unlike \cite{Nakai:2020cfw},  that 
we can look into this regime of $p$
is that the dark photon is massless during inflation and acquires mass, which can be very large,  only at the end of inflation through the dynamical symmetry breaking mechanism.


\section{Different limits}

In the following we first consider two specific limits: $\xi=0$ and $p=0$ separately. Then we consider cases where both $\xi$ and $p$ can be nonzero. We see that in all cases we can obtain enough relic energy density of vector dark matter in a wide range of dark photon mass.

\subsection{Limit I: Kinetic conformal coupling}\label{sec-model-kinetic}

In this section, we consider the kinetic conformal coupling case which corresponds to
\begin{eqnarray}\label{f}
\f(\tau) = f_{\rm e} \left(\dfrac{\tau}{\tau_{\rm e}}\right)^{p}\,, \hspace{1cm}
\g(\tau) = 0 \,,
\end{eqnarray}
which is equivalent to the case of $\gamma=0$ (or  $\xi=0$) in our analysis in the previous section. To get a non-vanishing range for tachyonic momenta we must have $p\geq1$.  For the vector dark matter scenario with which we are interested in this paper, the model Eq.~\eqref{f} is recently studied in Refs. \cite{Nakayama:2019rhg,Nakayama:2020rka,Nakai:2020cfw}. However, here we have a different and novel  scenario of  vector dark matter, thanks to the symmetry breaking mechanism which allows us to have much heavier dark photons.

\fg{
	\centering
	\includegraphics[width=0.45\textwidth]{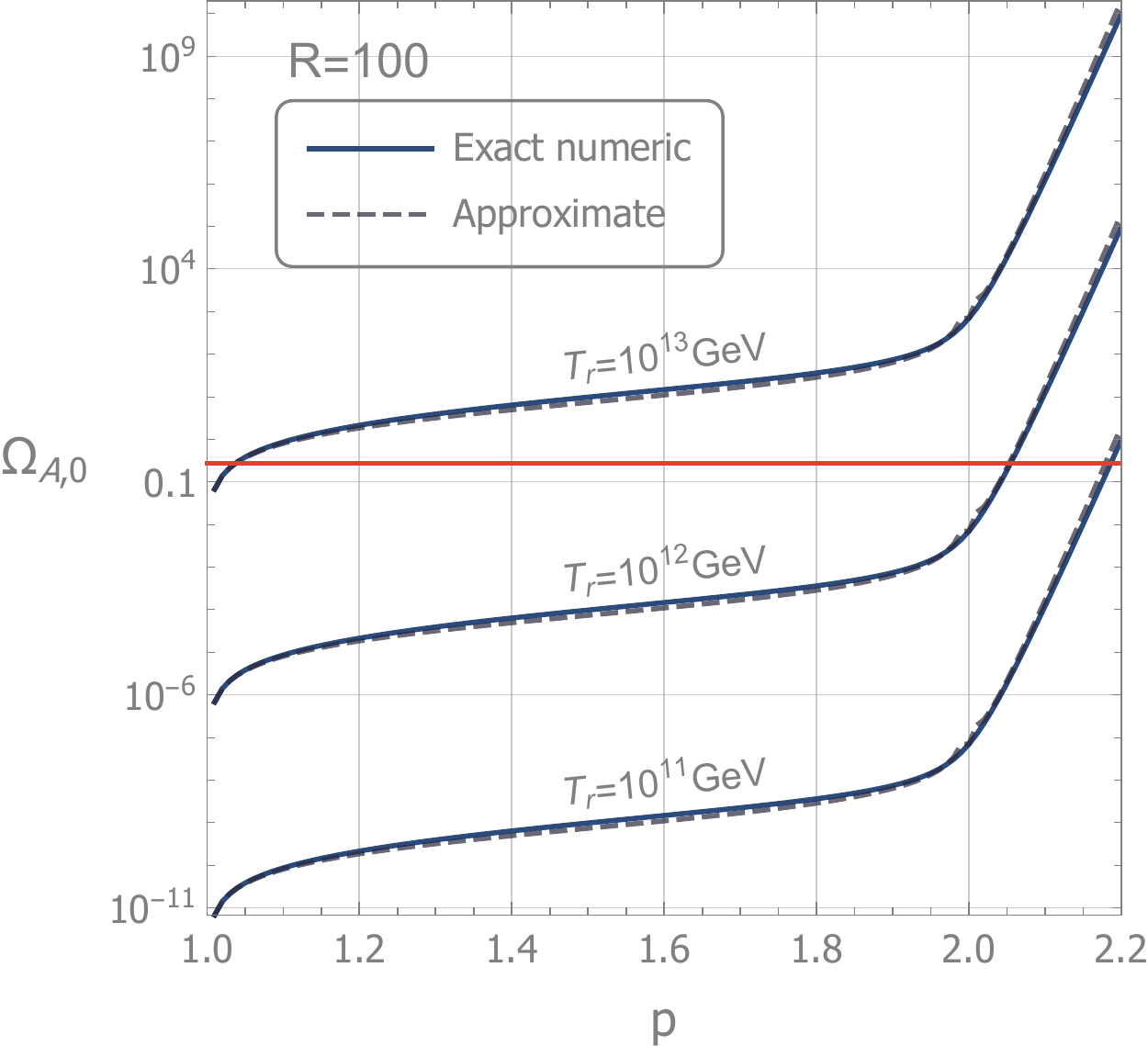}
	\hspace{8mm}
	\includegraphics[width=0.445\textwidth]{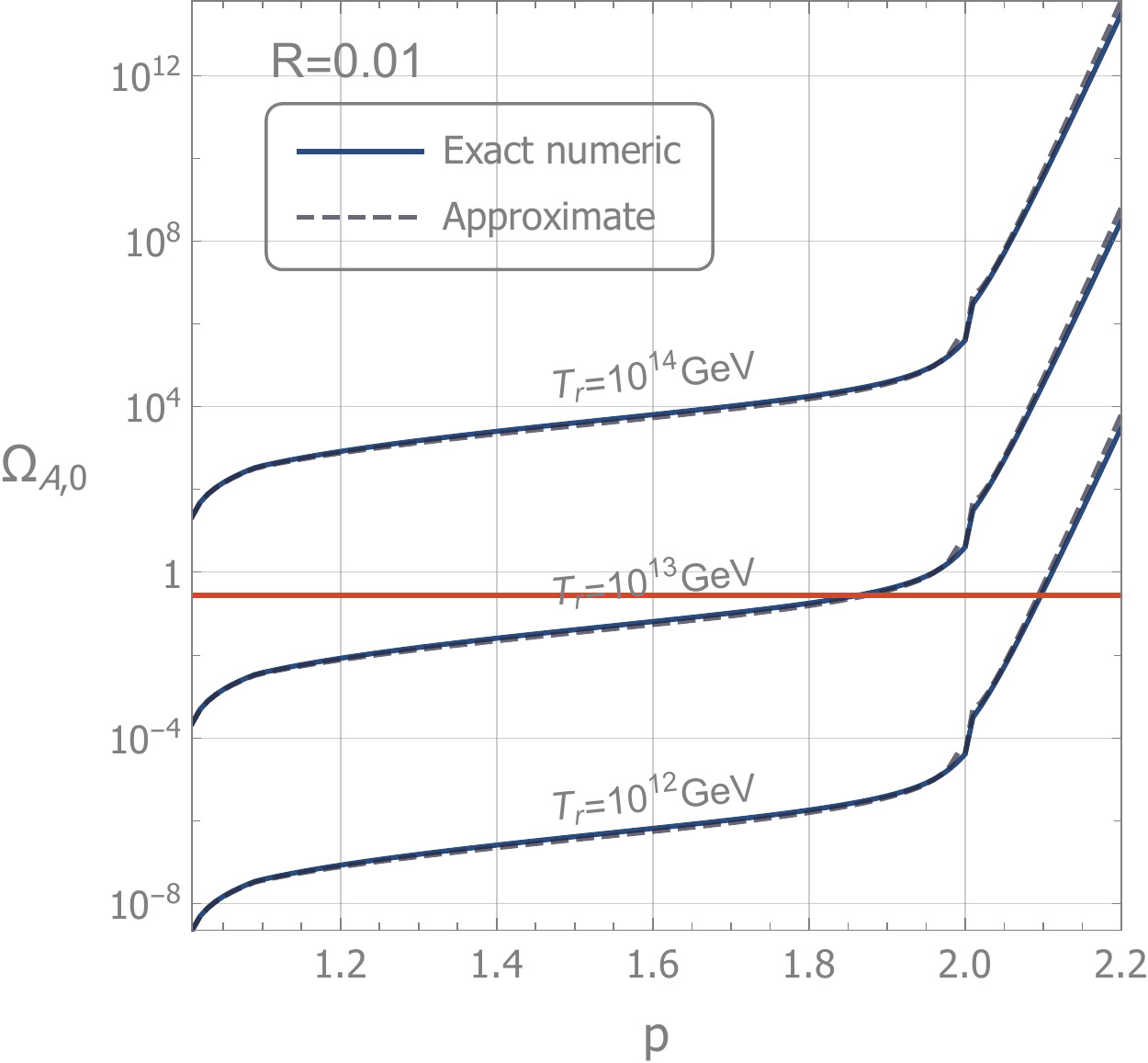}
	\caption{The present energy density of dark photons for $R=100$ (left) and $R=0.01$ (right). The numbers over the plot shows the reheating temperature $T_{\rm r}$. The red line represents the observed dark matter density $\Omega_{\rm DM}=0.27$. }
	\label{fig:relic1}
}

For $\xi=0$, using the identity $W(0,\nu,z) = \frac{\sqrt{\pi{z}}}{2} i^{\nu+1}H^{(1)}_{\nu}\left(\frac{iz}{2}\right)$ in which $H^{(1)}_\nu (x)$ is the Hankel function of the first kind, the mode function \eqref{mode} can be written in a  more familiar form 
\ba
\label{mode1}
v_k (\tau) = \frac{\sqrt{-\pi \tau}}{2} e^{i (1+ 2 \nu)\pi/4} H_\nu^{(1)} (- k \tau) \, ,
\ea
where as before $\nu=p-1/2$. Correspondingly, the expression in Eq.~\eqref{F} can also be written more compactly as
\eq{
	\label{rhoT2}
{\cal C}(p,\mathcal{N})=\frac{1}{8\pi}\int_{x_{\rm min}}^{x_{\rm max}}\dd{x} x^2 \Big[x^2
	\big|{ H_{\nu+1}^{(1)}(x)}\big|^2+\Big(x^2+\frac{R}{f^2}\Big)
	\big|{ H_{\nu}^{(1)} (x)}\big|^2\Big]\,,
}
where the integration region according to Eq.~\eqref{intreg1} is now given by
\eq{
	\label{intreg1-p}
	x_{\rm min}= \sqrt{p(p-1)} e^{-\mathcal{N}}\,,\hspace{1cm} x_{\rm max}=\sqrt{p(p-1)}\,,\qquad\,\, (p\geq1) \,.
}  
Note that by setting $\xi=0$ all dependences on polarizations disappear, i.e. both polarizations contribute equally to the dark matter energy density.

In the case with $p>2$, the dominant contribution to the integral in Eq.~\eqref{rhoT2} comes from the lower limit $x_{\rm min}$ ($\ll1$) and therefore we can use the small argument limit of the Hankel functions to the first order of approximation. On the other hand, for $1\leq p\leq 2$, it is not clear if we can use the small argument limit of the Hankel functions. However, we have checked numerically that in computing the integral in Eq.~\eqref{rhoT2} for all $p\geq1$ it is sufficient to use the small argument approximation of the Hankel function
\eq{
	\label{hankel}
|	H_{\nu}^{(1)} (x)|\simeq\frac{\Gamma(\abs{\nu})}{\pi}\left(\frac{x}{2}\right)^{-\abs{\nu}}\,.
}
Using the above approximation in Eq.~\eqref{rhoT2} and then performing the integration, we find the following analytic result
\eq{
\label{pg2}
{\cal C} (p,\mathcal{N})\simeq
\frac{2^{2p}\Gamma\left(p+\frac{1}{2}\right)^2}{8\pi^3|p-2|(p(p-1))^{p-2}}
\begin{cases}
\left(1+\frac{R}{(2p-1)^2}\right)e^{(2p-4)\mathcal{N}}\,,\qquad &(p>2)\\
2|p-2|(p(p-1))^{p-2}\left(1+\frac{R}{9}\right)\mathcal{N}\,,\qquad &(p=2)\\
\left(1+\frac{R}{(2p-1)^2}+\frac{p(p-1)(2-p)}{(3-p)(2p-1)^2}\right)\,, \qquad &(1<p<2)
\end{cases}\,.
}
We can use this to compute the relic density from Eq.~\eqref{relic-DMfinal}. We also need to compute the $t_{\rm NR}$ which depends on $R$ and $\kappa=\sqrt{p(p-1)}$. For instance, assuming $R>1$ the system becomes non-relativistic right at the end of inflation. However, for $R<1$ it will take some time for the system to become non-relativistic which depends on the minimum value of $R^{1/4}$ and $R^{1/2}/\kappa$.

Fig.~\ref{fig:relic1} shows the relic energy density today in this limit for $1\leq p\lesssim2.2$ (note that from the isocurvature bound we must have $p\lesssim2.01$) for two cases $R=100$ and $R=0.01$. The red horizontal line is the observed dark matter density today. We can see that assuming $10^{12}\lesssim T_{\rm r}\lesssim10^{13}$GeV we can get all of the dark matter in form of the vector dark matter with appropriate choice of $p$. Also the dashed curve in Fig.~\ref{fig:relic1} is obtained from the approximate analytic form of Eq.\eqref{pg2}. As a last comment note that the case $p<2$ is absent in the analysis of Ref. \cite{Nakai:2020cfw} since in their setup with an exponentially small value of $R$ ($R \sim 10^{-80} $ ), the produced dark photons would have very large momenta and will never become non-relativistic. In our setup with $R>1$, larger momenta can become non-relativistic and we can have vector dark matter even for the case $p<2$.

\subsection{Limit II: Chiral coupling}\label{sec-model-CS}

In this section we consider the case known as the pseudoscalar inflation with chiral interaction with the vector field and is defined by
\begin{eqnarray}\label{h}
\f(\tau) = 1\,, \hspace{1cm}
\g(\tau) = \gamma \f_{\rm e}\Big(\dfrac{\tau}{\tau_{\rm e}}\Big)^{q} \,,
\end{eqnarray}
which corresponds to the case with $p=0$. Here we add the assumption that $|q|\ll1$ such that $\xi$ defined in Eq.~\eqref{v-eq1} is approximately constant in time. Further, note that this model is very similar to axion-like coupling of inflaton to the gauge field in which 
\eq{
\label{sorbo}
\g^2(\phi)=\frac{\phi}{f_a}\,,
} 
where $f_a$ is a dimensionful parameter which determines the strength of the interaction (the axion decay constant).  In this case we can define \cite{Sorbo:2011rz}
\eq{
\xi \equiv \frac{\dot{\phi}}{2f_aH}\,,
}
which is approximately constant up to slow-roll parameters. With this definition, our phenomenological model also includes the results of the axion-like coupling case. Indeed, vector dark matter in the latter model with the coupling given in  Eq.~\eqref{sorbo} was recently studied in Ref. \cite{Bastero-Gil:2018uel}. However, in our symmetry breaking scenario we can have much heavier dark photons which makes our results different than those of Ref. \cite{Bastero-Gil:2018uel}. In Ref. \cite{Bastero-Gil:2018uel}, the authors also took into account the effects of time dependence of $\xi$ by numerical method. Here, we assume that $\xi$ is constant as even with this assumption we have a new scenario thanks to the symmetry breaking mechanism.

\fg{
	\centering
	\includegraphics[width=0.45\textwidth]{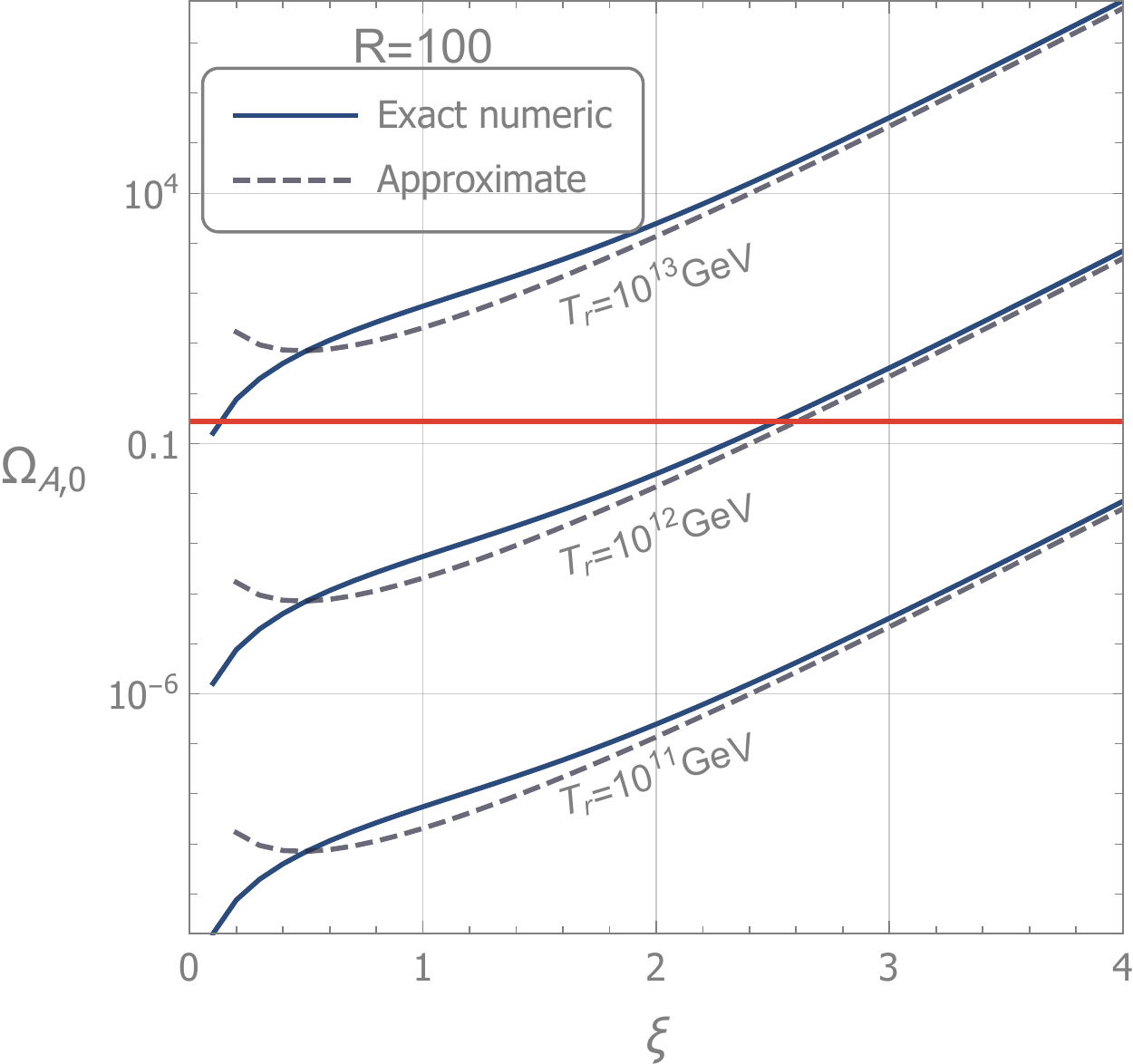}
	\hspace{8mm}
	\includegraphics[width=0.455\textwidth]{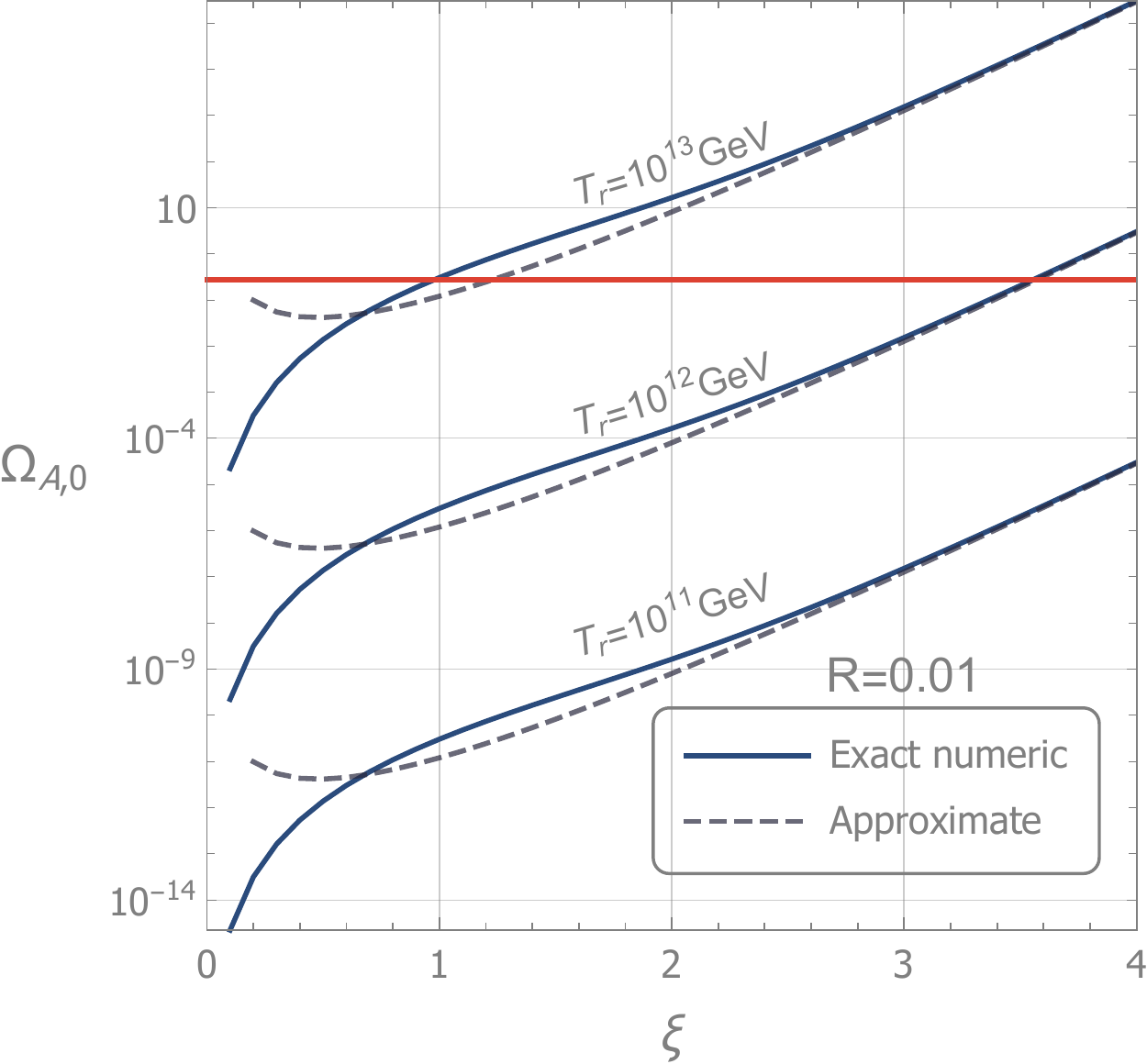}
	\caption{The present energy density of dark photons for both cases $R>1$ and $R<1$. The number over each plot shows the reheating temperature $T_{\rm r}$. The red line is the observed dark matter density $\Omega_{\rm DM}=0.27$.}
	\label{fig:relic1_axion}
}

Getting back to the model described by  Eq.~\eqref{h}, we can use our solution Eq.~\eqref{mode} by simply setting $p=0$. In this case we can use the identity $W_{\mu,1/2}(z)=e^{i\rm{arg}[\Gamma(1-\mu)]}e^{-i\pi\mu/2}\big(iF_0(i\mu,iz/2)+G_0(i\mu,iz/2)\big)$ in which $F_0$ and $G_0$ are the regular and irregular Coulomb functions respectively, to find
\ba
\label{mode-chiral}
v_{k,\lambda} (\tau) = \frac{e^{i\Theta}}{\sqrt{2k}} 
\Big( iF_0(\lambda\xi,-k\tau) + G_0(\lambda\xi,-k\tau) \Big)\,,
\ea
where $\Theta\equiv{\rm arg}[\Gamma(1+i\lambda\xi)]$ determines a constant phase. The above result coincides with the result of Ref. \cite{Sorbo:2011rz} up to the phase factor $e^{i\Theta}$ which is not important for our purpose. 

Contrary to the case of kinetic conformal coupling where we could use the small argument relation \eqref{hankel}, here we cannot expand the Coulomb functions to find an analytic expression for the energy density of the dark photons. However, as it is shown in Ref. \cite{Sorbo:2011rz}, we can instead use the WKB approximation to find
\eq{
v_k\approx\frac{1}{\sqrt{2k}}\Big(\frac{x}{2\xi}\Big)^{1/4}\exp(\pi\xi-2\sqrt{2\xi {x}})\,,
}
where we have assumed that $\xi>0$ as in previous section. As we see, the produced dark photons are completely chiral dominated by the plus polarization (with our convention with $\xi>0$). Following Ref. \cite{Anber:2006xt} and  using the approximate WKB solution for the mode function, we find that 
\eq{\label{F-chiral}
{\cal C} (\xi) \simeq \frac{6!}{2^{19}\pi^2} \Big(1+\frac{8}{15}R\Big) \frac{e^{2\pi\xi}}{\xi^3}\,,
}
where we have neglected the contribution from the minus polarization as it is completely negligible in comparison to the exponentially enhanced plus polarization.

Having Eq. \eqref{F-chiral} at hand, we can compute the relic energy density of dark matter from Eq.~\eqref{relic-DMfinal}. As in the previous section, we consider the cases $R>1$ and  $R<1$ separately. For  $R>1$ the vector field is non-relativistic right after inflation while for $R<1$ it takes some time for the energy density to become non-relativistic which depends on the value of $R$ and $\kappa=2\xi$. In both cases the relic energy density is given by the general expression \eqref{relic-DMfinal}. The relic density for dark matter versus the chirality parameter $\xi$ for both $R>1$ and $R<1$ are plotted in Fig.~\ref{fig:relic1_axion}. The solid curves are computed numerically from Eq.~\eqref{F} while the dashed curves are  the approximate analytic result Eq.  \eqref{F-chiral}. We can see a reasonable consistency between the numeric results except for small $\xi$ for which the WKB approximation breaks down. Finally, note that with $10^{12}\lesssim T_{\rm r}\lesssim10^{13}$GeV we can generate all of the observed dark matter from this model. 

\subsection{General case}\label{sec-model-general}

After considering two special cases of $\xi=0$ and $p=0$ in subsections \ref{sec-model-kinetic} and \ref{sec-model-CS}, in this subsection we consider the more general case where both $\xi$ and $p$ are nonzero. In this case we have an analytic solution for the mode function Eq.~\eqref{mode} in terms of the Whittaker function. However, we cannot perform the integrals in Eqs.~\eqref{F} and \eqref{G} analytically so we have to  perform the integrals numerically. The models studied in \ref{sec-model-kinetic} and \ref{sec-model-CS} then can be thought of as special cases of this general model. From Eq.~\eqref{relic-DMfinal}, for each value of $(p,\xi)$ we are interested in the corresponding reheating temperature $T_{\rm r}$ such that all observed dark matter abundance can be generated  in our model. Here, we set $R=100$ such that the dark photons become non-relativistic right after inflation. Other values of $R$ can be treated similarly taking into account the effect of $t_{\rm NR}$. 
 
The kinetic-like scenario for which $p\geq1$ is shown in the left panel of Fig.~\ref{fig:relicgeneral1} where the number over each contour is the logarithm of the corresponding reheating temperature in GeV units. We can see that a wide range of viable parameters can be used with temperature between around $10^{12}\sim10^{13}$GeV. Also, as we have discussed earlier, the net chirality of dark photons is controlled by the parameter $\xi$. In the right panel of Fig.~\ref{fig:relicgeneral1} we have plotted the parity parameter
\eq{
\label{delta}
\delta\equiv\frac{\rho_+-\rho_-}{\rho_++\rho_-}\,,
}
where $\rho_\pm$ is the relic energy density of the plus and minus polarizations respectively. One can see from the right panel of Fig.~\ref{fig:relicgeneral1} that upon increasing $\xi$ the plus polarization dominates very easily. 

\fg{
	\centering
	\includegraphics[width=0.45\textwidth]{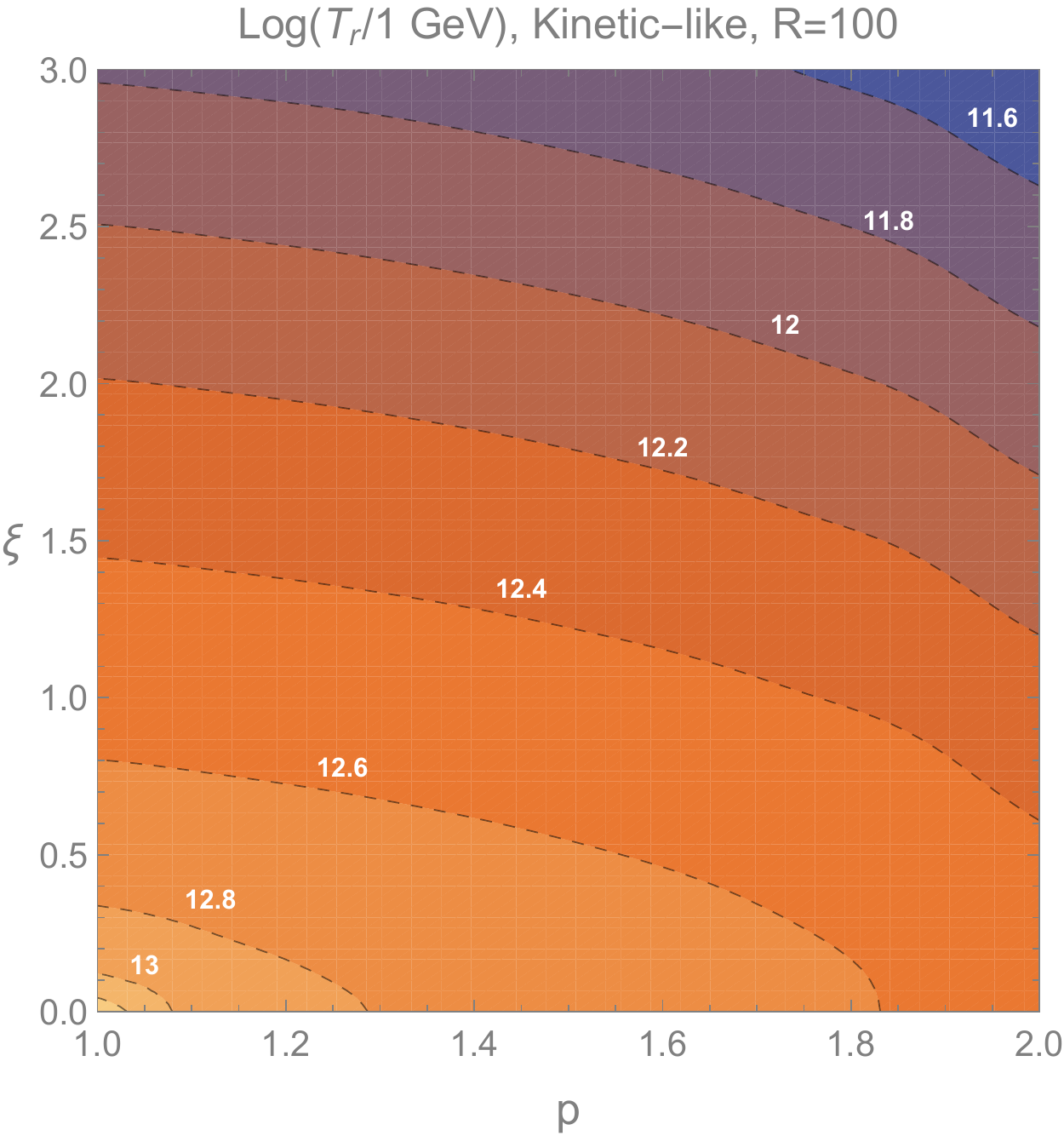}
	\includegraphics[width=0.53\textwidth]{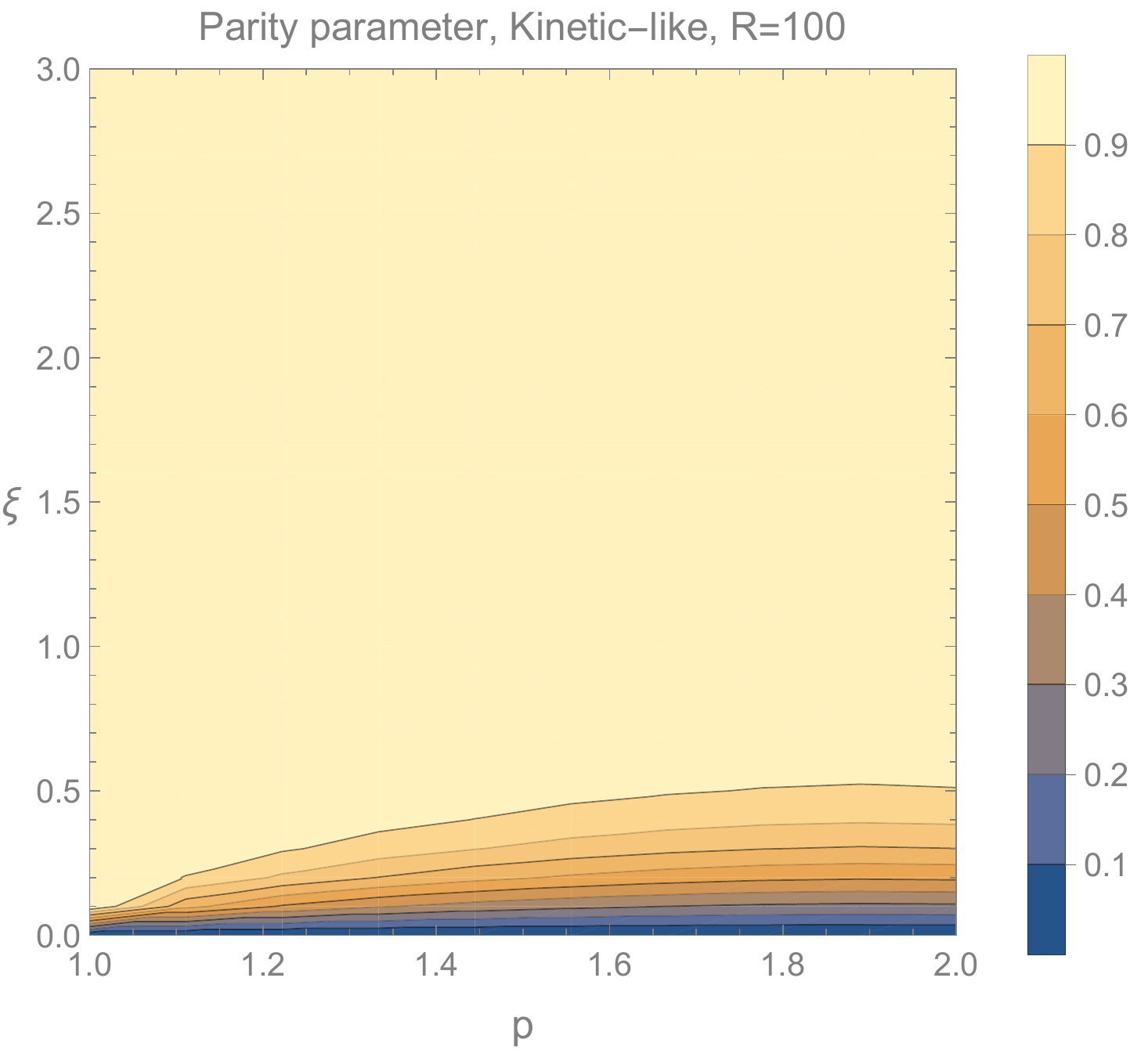}
	\caption{Kinetic-like case with $1<p<2$ and $R=100$. 
	Left: Logarithm  of the reheating temperature, $\log_{10}(T_{\rm r}/{\rm {GeV}})$, 
	needed to generate  the observed dark matter density $\Omega_{\rm DM}=0.27$. Right: The net chirality of the dark photons.}
	\label{fig:relicgeneral1}
}

Finally, the chiral-like scenario is shown in Fig.~\ref{fig:relicgeneral2} for which, as the other case, we have plotted the logarithm of the  reheating temperature in ${\rm GeV} $
unit. Once again, all dark matter can be obtained from our model assuming temperature between around $10^{12} - 10^{13}$GeV. Also in this case only the plus polarization is excited so the parameter $\delta$ defined in \eqref{delta} is nearly equal to unity. However, note that  there is a non-tachyonic region associated with the small values of  
$\xi$ as denoted by the  grey area in Fig.~\ref{fig:relicgeneral2}.

\fg{
	\centering
	\includegraphics[width=0.45\textwidth]{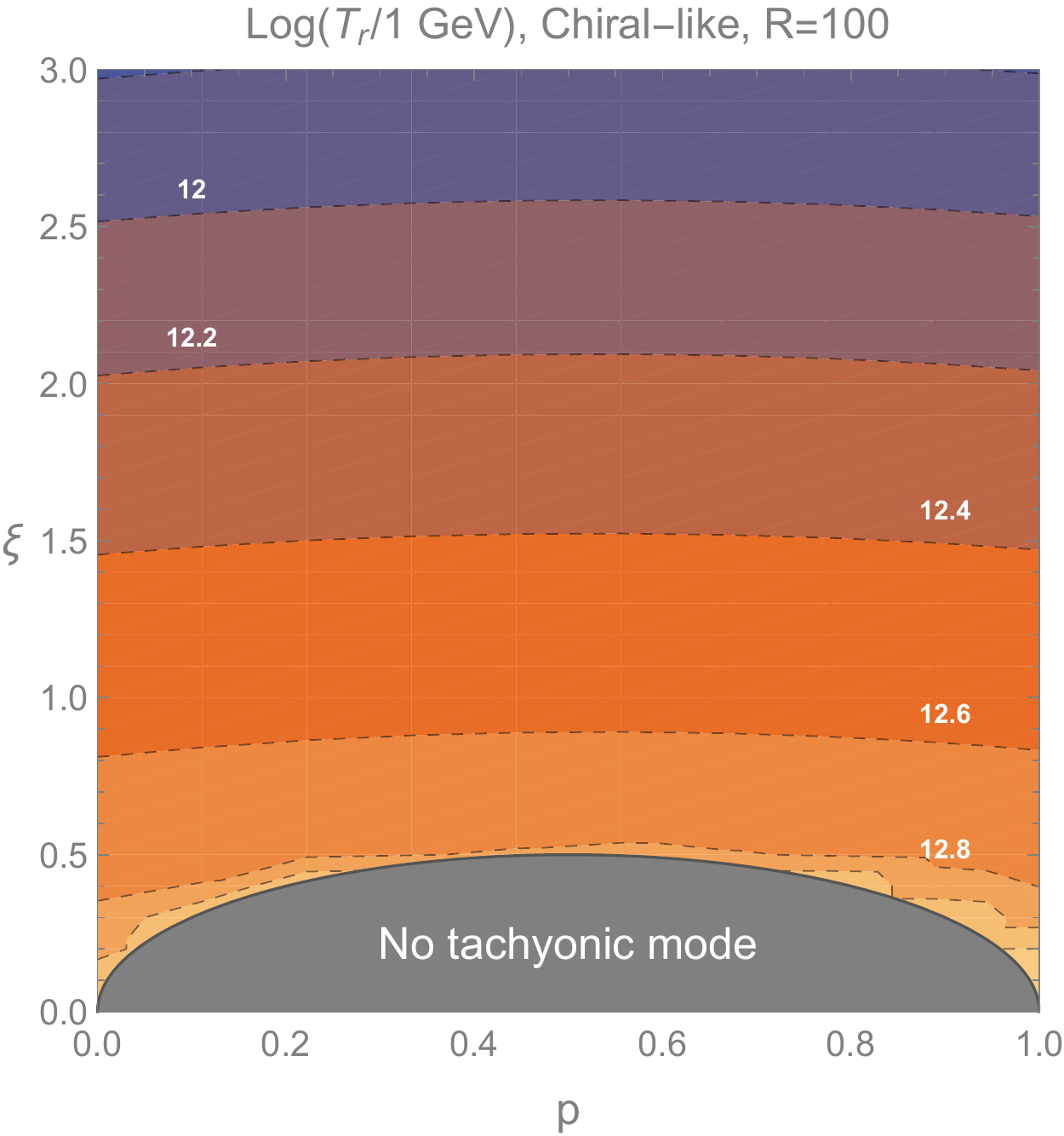}
	\caption{Chiral-like case with $0<p<1$ and $R=100$. We have plotted the 
	logarithm of  the reheating temperature, $\log_{10}(T_{\rm r}/{\rm {GeV}})$,  
	 needed to generate  the observed dark matter density $\Omega_{\rm DM}=0.27$. }
	\label{fig:relicgeneral2}
}

\section{Summary and conclusions}\label{summary}

The search for the nature of dark matter is an active area of research both in particle physics and in cosmology. The popular models of dark matter are mostly in the context of beyond SM particle physics such as the WIMPs scenarios which are still waiting for the direct or indirect verifications. On the other hand, inflation is a cornerstone of early universe cosmology, a working mechanism to generate large scale structure in the observed universe. It is an interesting question  if dark matter can be produced naturally in the context of inflationary dynamics. In this work we have revisited the idea of ``vector dark matter'' in the context of inflation with symmetry breaking. 
  
It has been shown that the longitudinal mode of dark gauge boson during inflation can play the roles of vector dark matter after inflation \cite{Graham:2015rva}. Recently, it was shown to be even possible to obtain the correct relic density for the vector dark matter from the transverse modes \cite{Bastero-Gil:2018uel,Nakayama:2019rhg,Nakayama:2020rka,Nakai:2020cfw}. The idea is to pump energy from the inflaton to the dark photons sector through some appropriate couplings to prevent the gauge field energy density from being diluted during inflation.  In Ref. \cite{Bastero-Gil:2018uel} the model with the parity violating interaction term  $\phi F \tilde F$ has been studied and 
it is shown that to have the correct dark matter relic abundance  the  dark photon's mass is at the order of $\m \geq 10^{-6}$ eV. On the other hand, in Ref. \cite{Nakai:2020cfw} the authors considered the kinetic conformal coupling model and have found much lighter mass for the dark photons $\m \sim 10^{-21}$ eV. The difficulty with these scenarios is that on one hand one needs to assume a small mass during inflation to have an efficient dark photon production. On the other hand, one needs that the dark photon's mass to be large enough so that the energy density of the dark photon becomes non-relativistic before the time of matter-radiation equality. 

In this paper, we have presented a new vector dark mater scenario which relaxes these difficulties. In this model, the gauge field is massless during almost all period of inflation and therefore the dark photons can be  produced efficiently. Then the gauge field acquires mass through a dynamical symmetry breaking mechanism toward the end of inflation. This dynamical mass generation mechanism for the dark photons allows us to have large mass right after inflation. In this respect, we have significantly enlarged the parameter space of the models considered in Refs. \cite{Bastero-Gil:2018uel} and \cite{Nakayama:2019rhg,Nakayama:2020rka,Nakai:2020cfw} by relaxing the condition of having a small mass. Moreover, we have investigated a new setup containing both models of Refs. \cite{Bastero-Gil:2018uel} and \cite{Nakayama:2019rhg,Nakayama:2020rka,Nakai:2020cfw} simultaneously. We have shown that in this general model, we can obtain the correct relic abundance of the vector dark matter for a large space of the model parameters.

The symmetry breaking at the end of inflation allows us to have a dark photon much heavier than or comparable to the Hubble scale at the end of inflation, corresponding respectively to $R\gg1$ and $R\sim1$.   
With $R$ not exponentially smaller than unity our setup  predicts that we need a reheat temperature around and lower than $10^{12} {\rm GeV}$ in order to generate  the observed dark matter abundance today. This is not a very high reheat temperature. As a result, our setup predicts that the tensor to scale ratio parameter $r$ to be very small, typically $r \lesssim 10^{-14}$, which is too small to be detected in any foreseeable cosmological observations. Of course, choosing an exponentially small value of $R$, our model can saturate  the current upper bound $r \lesssim 10^{-2}$. 

With an instantaneous reheating scenario the mass of dark photon is related
to the reheat temperature via $m_A \sim \sqrt R T_{\rm r}^2/M_P$. So with $T_{\rm r} =10^{12} {\rm GeV}$
we obtain $m_A \sim \sqrt R \times 10^6 {\rm GeV}$. For example with $R=10^2$ we obtain $m_A \sim 10^7{\rm GeV}$ while for $R=10^{-2}$ we have $m_A \sim 10^5{\rm GeV}$.  As a result,  the dark photon in our setup is  typically much heavier  compared to the conventional WIMPs scenarios. However, with an exponentially small value of $R$,  much lighter dark photons can also be generated as well. Another generic prediction of our setup is that cosmic strings are produced at the end of inflation. Using the CMB upper bound on the tension of strings we have obtained 
the approximate upper bound on $R$ in Eq. (\ref{R-bound}). On the other hand, demanding that dark photon becomes non-relativistic before the time of matter and radiation equality, we find Eq. 
\eqref{R-cond1} which gives a lower bound on $R$. From Eqs. (\ref{R-bound}) and (\ref{R-cond1}) we find a wide mass range of $\m \in [10^{-20}$eV$-10^{15}$GeV$]$ for dark photon. We note that possible decay of dark photon to the SM particles through its interaction with the inflaton and/or standard radiation would narrow this mass range while still we can have dark photon with the large mass $\m \geq$ GeV in our model.  

In this work we have employed a heavy complex field, like the waterfall field of hybrid inflation, to undergo symmetry breaking and generate the mass for the dark gauge field via the Higgs mechanism. However, it is possible to consider a scenario in which the inflaton field itself is complex and is charged under the dark gauge field. If the inflationary potential has global minima as in the Higgs potential, then the mechanism of vector dark matter  production may occur in this setup as well. We would like to come back to this question elsewhere.

Finally we comment that to simplify the analysis we have neglected the dark photon production during the (p)reheating stages. Specifically, because of the couplings $f(\phi)^2$ and $h(\phi)^2$ the quanta of dark photons can in principle be generated via parametric resonance. To avoid this complication we have assumed that $\phi-$dependence in the above couplings become very weak after inflation. Otherwise, one must take into account the additional energy transfer from the oscillating inflaton field to the dark photons during preheating as well. This can be particularly important for the dark photons lighter than $H_{\rm e}$, i.e. in models with $R <1$. However, for heavy dark photons with  $R>1$ the particle production is not efficient and one can safely neglect its contribution in the observed dark matter relic density. Moreover, the details of the efficiency of the dark photon production via parametric resonance depend on the shape of the inflaton potential around its minimum.  This is an interesting question which is beyond the scope of this work but we would like to come back to this question in future.

\vspace{1cm}

{\bf Acknowledgments:} 
We would like to thank A. A. Abolhasani, S. Khosravi and M. H. Namjoo for insightful comments. We also thank the anonymous referee for very helpful comments. The work of M.A.G. was supported by Japan Society for the Promotion of Science Grants-in-Aid for international research fellow No. 19F19313. The work of S.M. was supported in part by Japan Society for the Promotion of Science Grants-in-Aid for Scientific Research No.~17H02890, No.~17H06359, and by World Premier International Research Center Initiative, MEXT, Japan. 

\vspace{0.7cm}

\bibliography{ref} 
\bibliographystyle{JHEP}

\end{document}